\documentclass[manuscript]{aastex}
\usepackage{graphicx}
\usepackage{natbib}
\usepackage{amssymb}

\begin{document}

\newcommand{\Po}{\ion{P}}
\newcommand{\Ti}{\ion{Ti}}
\newcommand{\feii}{\ion{Fe}{2}}
\newcommand{\feiii}{\ion{Fe}{3}}
\newcommand{\h}{H$_2$}
\newcommand{\ci}{\ion{C}{1}}
\newcommand{\oi}{\ion{O}{1}}
\newcommand{\oii}{\ion{O}{2}}
\newcommand{\sii}{\ion{S}{2}}
\newcommand{\nii}{\ion{N}{2}}
\newcommand{\n}{\ion{N}{1}}
\newcommand{\he}{\ion{He}{1}}
\newcommand{\tiii}{\ion{Ti}{2}}
\newcommand{\crii}{\ion{Cr}{2}}
\newcommand{\caii}{\ion{Ca}{2}}
\newcommand{\pii}{\ion{P}{2}}
\newcommand{\hi}{\ion{H}}

\slugcomment{To appear in Ap. J.}

\title{The diagnostic potential of Fe lines applied to protostellar jets\thanks{Based on observations collected with X-shooter at the Very Large
Telescope on Cerro Paranal (Chile), operated by the European Southern
Observatory (ESO). Program ID: 085.C-0238(A).}}

\author{T. Giannini\altaffilmark{1}, B. Nisini\altaffilmark{1}, S. Antoniucci\altaffilmark{1}, J. M. Alcal\'a\altaffilmark{2}, F. Bacciotti\altaffilmark{3}, R. Bonito\altaffilmark{4,5}, L. Podio\altaffilmark{6}, B. Stelzer\altaffilmark{4}, E.T. Whelan\altaffilmark{7}
}
\altaffiltext{1}{INAF-Osservatorio Astronomico di Roma, via Frascati 33, I-00040 Monte Porzio Catone, Italy}
\altaffiltext{2}{INAF-Osservatorio Astronomico di Capodimonte, via Moiariello 16, I-80131 Napoli, Italy} 
\altaffiltext{3}{INAF-Osservatorio Astrofisico di Arcetri, Largo E. Fermi 5, I-50125 Firenze, Italy} 
\altaffiltext{4}{INAF-Osservatorio Astronomico di Palermo, Piazza del Parlamento 1, I-90134 Palermo, Italy}
\altaffiltext{5}{Dipartimento di Fisica e Chimica, Universit\'a di Palermo, Piazza del Parlamento 1, I-90134 Palermo, Italy}
\altaffiltext{6}{UJF-Grenoble 1 / CNRS-INSU, Institut de Planetologie et d'Astrophysique de Grenoble (IPAG) UMR 5274, Grenoble, F-38041, France}
\altaffiltext{7}{Institut f\"{u}r Astronomie und Astrophysik, Kepler Center for Astro and Particle Physics, Eberhard Karls Universit\"{a}t, 72076 T\"{u}bingen, Germany}
\begin{abstract}
We investigate the diagnostic capabilities of the iron lines {\bf for} tracing the physical conditions of the shock-excited gas in jets driven by
pre-main sequence stars. 
We have analyzed the 3\,000-25\,000 \AA\,, X-shooter spectra of two jets driven by the pre-main sequence stars ESO-H$\alpha$ 574 and Par-Lup 3-4. Both spectra 
are very rich in [\feii]  lines {\bf over} the whole spectral range; in addition, lines from 
[\feiii] are detected in the ESO-H$\alpha$ 574 spectrum. NLTE codes solving the equations of the statistical equilibrium along with codes for the ionization equilibrium are used to derive the gas excitation conditions of electron temperature and density, and fractional ionization. An estimate of the 
iron gas-phase abundance is provided by comparing the iron lines emissivity with that of neutral oxygen at 6300 \AA.
The [\feii] line analysis indicates that the jet driven by ESO-H$\alpha$ 574 is, on average, colder ($T_{\rm e}$ $\sim$ 9\,000 K), less dense ($n_{\rm e}$ $\sim$ 2 10$^4$
cm$^{-3}$) and more ionized  ($x_{\rm e}$ $\sim$ 0.7) than the Par-Lup 3-4 jet ($T_{\rm e}$ $\sim$ 13\,000 K, $n_{\rm e}$ $\sim$ 6 10$^4$ cm$^{-3}$, $x_{\rm e}$ $<$ 0.4), even if the
existence of a higher density component ($n_{\rm e}$ $\sim$ 2 10$^5$ cm$^{-3}$) is probed by the [\feiii] and [\feii] ultra-violet lines. The physical conditions derived from the iron lines are
compared with shock models suggesting that the  shock at work in ESO-H$\alpha$ 574 is faster and 
likely more energetic than the Par-Lup 3-4 shock. This latter feature is confirmed  by the high percentage of gas-phase iron measured in ESO-H$\alpha$ 574 (50-60\% of its solar abundance in comparison with less than 30\% in Par-Lup 3-4), which testifies that the ESO-H$\alpha$ 574 shock is powerful enough to partially destroy the dust present inside the jet.
This work demonstrates that a multiline Fe analysis can be effectively used to probe the excitation and ionization conditions of the gas in {\bf a jet} without any assumption on ionic abundances. The main limitation on the diagnostics resides in the large uncertainties of the atomic data, which, however, can be overcome through a statistical approach involving 
many lines.
\end{abstract}

\section{Introduction}{\label{sec:sec1}
Jets from young stars play a key role in the dynamics of star formation and disk
evolution. They regulate the process of stellar
accretion, by both removing the angular momentum generated by accreting material in the disk, and modifying
the inner disk physics, thus influencing the evolution of proto-planetary systems. 
The specific role of jets in the dynamics and evolution of the accreting system strongly depends on the parameters
that characterize their structure and excitation, which are in turn related to their formation and heating mechanisms.
From an observational point of view, information on the jet physics and dynamics can be retrieved through the analysis of the forbidden lines emitted
by the jet plasma when it gets excited in shocks: to this aim, strong optical lines, such as [\oi], [\sii] and [\nii] lines,
are widely used and specific diagnostic tools, able to retrieve a complete set of parameters (namely electron density, n$_{e}$, 
temperature, T$_{e}$ and ionization fraction, x$_{e}$) \,have been developed (e.g. Bacciotti \& Eisl\"{o}ffel 1999).
The knowledge of these parameters is fundamental to an understanding of jet
acceleration mechanisms (e.g. MHD disk-winds or X-winds, e.g., Shu et al. 1994; Ferreira 1997)
and for measuring the mass flux rate (\.{M}$_{\rm{jet}}$). \.{M}$_{\rm{jet}}$ is the quantity regulating the efficiency 
of the jet and {\bf is} directly related to the disk mass accretion rate (\.{M}$_{\rm{acc}}$). 

Although widely exploited and constantly refined,  the diagnostic tools based on bright optical lines 
suffer from several intrinsic limitations. Firstly, optical lines trace specific 
excitation conditions and hence force the assumption that the gas in the jet has a constant temperature and density. This assumption
is in contrast with combined optical/near infrared line analysis, which has shown that gradients in temperature and density up
to orders of magnitudes usually occur in the cooling region behind the shock front (Nisini et al. 2005, Podio et al. 2006).
Secondly, diagnostic tools based on ratios between lines of different atomic species require one to assume a set of elemental
abundances, which in turn imply an uncertainty on the parameters (temperature and density) more than 40\% (Podio et al. 2006). 
Finally, the use of optical lines requires an {\it a priori} knowledge of the visual extinction, a circumstance that
often makes the optical diagnostic applicable only to jets of the more evolved sources, where the reddening is negligible.
All the above limitations can be circumvented by using different lines of the same species, covering a wide range of wavelengths
particularly sensitive to extinction variations.
In this respect diagnostic of iron (in different ionization stages) represents a very well suited tool. Indeed, since iron line spectrum covers all the
wavelengths between the Ultraviolet (UV) and the Near-Infrared (NIR), it is sensitive to a large range of excitation conditions, allowing therefore to derive a complete
view of the post-shock cooling region.
The aim of the present paper is to probe the  potential of the iron lines in probing the jet physical parameters. Our test-cases are two jets 
we have observed with the X-shooter spectrograph in the wavelength range $\sim$ 3\,000$-$25\,000 \AA, namely the jets excited by the sources 
ESO-H$\alpha$ 574 and Par-Lup 3-4.\\

ESO-H$\alpha$ 574 ($\alpha_{J2000.0}$= 11$^h$ 16$^m$ 03$^s$.7, $\delta_{J2000.0}$= $-$76$^\circ$ 24$^{\prime}$ 53$^{\prime\prime}$), spectral type K8,  is a low-luminosity source in the Chamaleon I star-forming region located at a distance $d$=160$\pm$17 pc (Wichmann et al. 1998). The low luminosity  of 3.4 10$^{-3}$ $L_{\odot}$ (Luhman 2007), which is a factor $\sim$ 150 lower than the luminosity of the typical T Tauri stars of the same spectral type, is interpreted as due to a disk seen edge-on. The source powers a bipolar
jet (HH 872) of total projected length of 0.015 pc (3140 AU). It was discovered by  Comer\'{o}n \& Reipurth (2006) in a [\sii] image at 6728 \AA\,  as a chain of knots, of which knots A1, A, B, C, D form the blue-shifted jet  and knot E forms the red-shifted jet. 

Par-Lup 3-4  ($\alpha_{J2000.0}$= 16$^h$ 08$^m$ 51$^s$.44, $\delta_{J2000.0}$= $-$39$^\circ$ 05$^{\prime}$ 30$^{\prime\prime}$), spectral type M5, is located in the Lupus III dark cloud at $d$=200$\pm$40 pc (Comer\'{o}n et al. 2003). This object also appears to be under-luminous, being about 25 times fainter than typical M5 pre-main sequence objects (L=3 10$^{-3}$ L$_{\odot}$, Mer\'{i}n et al. 2008). As in the case of ESO-H$\alpha$ 574, its low luminosity is likely due to the obscuration of the star by an edge-on viewing disk (Hu{\'e}lamo et al. 2010).
The jet was discovered by Fern\'{a}ndez \& Comer\'{o}n (2005), with emission extending in opposite directions with respect to the star for a total length of $\sim$ 1240 AU.

The X-shooter spectra of the two objects have been already investigated by Bacciotti et al. (2011, hereinafter BWA11) and by Whelan et al. (2013, hereinafter WBA13). Both these papers determine the mass ejection to mass accretion ratio \.{M}$_{\rm{jet}}$/\.{M}$_{\rm{acc}}$.
While in Par-Lup 3-4 this ratio is at the upper end of the range predicted by jet models, the value found in  ESO-H$\alpha$ 574 of $\sim$\,90 can be partially reconciled with the predictions of magneto-centrifugal jet acceleration mechanisms only if the edge-on disk severely reduces the luminosity of the accretion tracers. The numerous spectral lines detected in the two jets, along with the kinematical properties derived in the line profiles are presented in WBA13.\\
In the present paper we concentrate our analysis on the many iron lines detected in both  spectra. The outline is the following: in Sect.\,2 we briefly summarize the details of the observations and present the spectroscopic data; in Sect.\,3 we describe the iron excitation and ionization models and derive the jet physical parameters. In Sect.\,4  we discuss the results, which are summarized in Sect. 5.

\section{Observations and Results}{\label{sec:sec2}
The present work is part of a coherent series of papers that deal with our X-shooter survey of Pre-Main Sequence (PMS) objects. The
overall aspects, such as scopes, data reduction procedures, calibrations and results are thoroughly discussed in Alcal\'a et al. 2011 and  Alcal\'a et al. 2013. Here we just recall the
information which is essential for the presented subject.
The X-shooter spectra of ESO-H$\alpha$ 574 and Par-Lup 3-4 were acquired on April 7 2010, with an integration time of $\sim$ 1 hr per object. The slit, aligned with the jet axis,
was set to achieve a resolving power of 5\,100, 8\,800 and 5\,600 for the UVB (3\,000 - 5\,900 \AA),\, VIS (5\,450 - 10\,200 \AA)\, and NIR arm (9\,900 - 24\,700 \AA)\,, respectively (slit widths: 1$\farcs$0,  0$\farcs$9, 0$\farcs$9). The pixel scale is 0$\farcs$16 for the UVB and VIS arms and 0$\farcs$21 for the NIR arm.\\
The data reduction was performed independently for each arm using the X-shooter pipeline version 1.1., which provides 2-dimensional spectra, background-subtracted and calibrated in wavelength. Post-pipeline procedures were then applied by using  routines within {\bf the} {\it IRAF} and {\it MIDAS} packages to subtract sky lines and obtain 1-dimensional spectra. These
are then divided by a telluric spectrum to remove the atmospheric features, and to do the flux-calibration}. The complete spectrum was obtained by comparing the flux densities in the overlapping portions of the spectra of adjacent arms. While UVB and VIS spectra are perfectly aligned, the NIR spectrum of ESO-H$\alpha$ 574 appears lower of a factor $\sim$ 1.26. Flux losses in the NIR arm are not uncommon in the X-Shooter spectra and are caused by  a misalignment between the NIR with respect to the VIS and UVB arms (Alcal\'a et al. 2013). No correction was need to re-align the three arms spectra of Par-Lup 3-4. \\   
As far as the ESO-H$\alpha$ 574 jet is concerned (hereinafter ESO-H$\alpha$ 574), we concentrate here on the iron lines detected in the brightest knot A1. This is also the closest to the exciting source, extending from the source itself up to 2$^{\prime\prime}$ away (320 AU, see Figure 2 of BWA11, upper panel). The Par-Lup 3-4 jet (hereinafter Par-Lup 3-4) was integrated up to a distance of 1$^{\prime\prime}$ from the source continuum, which corresponds to 200 AU (see Figure 2 of BWA11, lower panel).\\
Figures \ref{fig:spettro_uv} and \ref{fig:spettro_visnir} show the portions of the spectra of the two objects where iron lines are detected, while 
Figure\,\ref{fig:livfeII} shows the Grotrian diagram of Fe$^+$ levels from which the detected lines originate. The maximum energy level is at more than 30\,000 cm$^{-1}$ above the ground state, and the line wavelengths  cover the whole investigated range (in blue, green, and red we indicate ultra-violet, optical, and near-infrared lines, respectively). Similarly, Figure\,\ref{fig:livfeIII} gives the diagram of Fe$^{++}$ levels. Note that, due to the level structure, all the emitted lines lie only in the ultra-violet range,
although the covered energy range is comparable to that of Fe$^{+}$.\\
The line fluxes of all the detected lines are listed in Table A.1 of WBA13. Here we give in Table\,\ref{tab:tab1} the observed line ratios R$_{ESO-H\alpha574}$ and R$_{Par-Lup3-4}$ of the [\feii] lines detected in the two objects with respect to the bright line at 4277 \AA. Lines originating from the same multiplet are grouped together and listed in order of decreasing energy of the upper level. In the last column, the ratio R$_{ESO-H\alpha574}$/R$_{Par-Lup3-4}$ is reported. Since the differential extinction between the two objects is negligible (see Sect. \ref{sec:sec3.1.1}), this ratio gives a qualitative indication on whether or not the excitation conditions are similar in the two objects. Indeed,
lines with excitation energy $\ga$ 20\,000 cm$^{-1}$ (ultra-violet and optical lines) and those with excitation energy $\la$ 20\,000 cm$^{-1}$ (near-infrared lines) have $<R_{ESO-H\alpha574}$/R$_{Par-Lup3-4}> \approx$ 1.15 and 
$<R_{ESO-H\alpha574}$/R$_{Par-Lup3-4}> \approx$ 2.6, respectively. This in practice suggests that in Par-Lup 3-4 the most excited lines are brighter (in comparison to the 4277 \AA) than in ESO-H$\alpha$ 574, a circumstance that could reflect a higher gas temperature.\\
Notably, [\feiii] lines are detected only in ESO-H$\alpha$ 574 (see Table\,\ref{tab:tab2}). This result cannot be explained with a different sensitivity in the X-shooter spectra of the two jets, which are similarly bright and were integrated for a comparable amount of time. Therefore, this feature also points to different excitation conditions in the two objects.

\section{Line fitting model}{\label{sec:sec3}
In this section we describe the excitation and ionization models which provide the main physical parameters of the two investigated jets. The results of the comparison between observations and models are summarized in Table\,\ref{tab:tab4}.

\subsection{The excitation model}{\label{sec:sec3.1}
The observed ratios between lines from the same ionic species (e.g. Fe$^{+}$ or Fe$^{++}$)
can be compared with the predictions by an excitation model to derive the physical conditions of the gas. To this aim we adopted a Non Local Thermal Equilibrium (NLTE) approximation for line excitation. One of the main issues of such line modeling regards the choice of the atomic dataset.
The complexity of the iron
atomic system, which involves hundreds of energy levels (with multiple metastable levels), makes it very difficult to get accurate atomic data sets (both radiative and collisional).
For example, seven different computations of the Einstein coefficients for the spontaneous radiative decay (A-values) have been implemented for Fe$^{+}$, which may differ from each other by more than 50\%. Bautista et al. (2013) have evaluated
the uncertainties in the line emissivities due to the combinations of the uncertainties on A-values, collisional coefficients and propagation of these two on the level populations.
For typical shock-excitation conditions, namely $T_{\rm{e}}$ $\sim$ 10\,000 K and density between 10$^2$-10$^8$ cm$^{-3}$, they find a very wide range of uncertainties, which vary from 
less than 10\% (e.g. lines at 1.256 $\mu$m and 8616 \AA) to more than 60\% (e.g. lines at 5.330 $\mu$m and at 5527 \AA). As shown by the same authors, the most effective way to circumvent the problem is to apply a statistical approach by including in the analysis a large number of lines.\\
In our model we use the up-to-date atomic database of the XSTAR compilation (Bautista \& Kallman 2001\footnote{available at heasarc.gsfc.nasa.gov/xstar/xstar.html}), which
gives energy levels, A-values and rates for collisions with electrons (these latter for temperatures between 2\,000 K and 20\,000 K) for the first 159 and 34 fine-structure levels of Fe$^+$ and Fe$^{++}$, respectively. The implications on the results when adopting different data sets will be commented in Sect.\ref{sec:sec3.1.1}. 

The NLTE model assumes  electronic collisional excitation/de-excitation and spontaneous radiative decay. Possible contributions on line emissivities due to radiative processes are discarded at this step of the analysis but will be considered in Sect.\,\ref{sec:sec3.3}.  
The free parameters of the excitation model are the electron temperature $T_{\rm{e}}$ and the density  $n_{\rm{e}}$, which can be derived from the observed flux ratios once these latter are corrected for the visual extinction ($A_{\rm{V}}$) along the line of sight. This latter parameter is usually derived from the flux ratio of lines emitted from the same upper level, this being independent from the level population and therefore a function only of the line frequencies and the A-coefficients. As stated above, however, the large uncertainties associated with these latter values, are reflected in a poor estimate of $A_{\rm{V}}$, especially if one considers only two or three lines, as is often done with the NIR [\feii] lines (see also Giannini et al. 2008). Therefore, we have taken the extinction as a further free parameter of the excitation model. To derive the differential extinction at each line wavelength, we adopt the extinction curve by Draine (2003).
To minimize the uncertainties, we included in the fit only the un-blended lines detected with a signal-to-noise (snr) ratio larger than 5 (i.e. 35 lines for ESO-H$\alpha$ 574 and 20 lines for Par-Lup 3-4)
and checked the compatibility of the fit with line fluxes at lower snr ratio {\it a posteriori}. First, we constructed a grid of model solutions in the parameter space 2\,000 K $<
 T_{\rm{e}} <$ 30\,000 K (in steps of $\delta T_{\rm{e}}$\,=\,1\,000 K); 10$^2$ cm$^{-3} < n_{\rm{e}} < $10$^7$ cm$^{-3}$ (in steps of log$_{10} (\delta n_{\rm{e}}/cm^{-3}$)\,=\,0.1) and $A_{\rm{V}}$ $\le$ 2 mag (in steps of $\delta A_{\rm{V}}$\,=\,0.5 mag). Then, following the method for line fitting proposed by Hartigan \& Morse (2007), we have iteratively changed the line used for the normalization, hence considering all the possible sets of line ratios. Each of them was then compared with the grid of theoretical values to find the model with the lowest value of $\chi^2$.

\subsubsection{[\feii] lines fit} {\label{sec:sec3.1.1}
The result of the excitation model considering the complete set of [\feii] lines detected in  ESO-H$\alpha$ 574 is depicted in Figure\,\ref{fig:bestfit_eso}. The minimum $\chi^2$-value is found if  the line at 4277 \AA \, is taken as a reference and the corresponding line ratios are reported in Table\,\ref{tab:tab1}. 
The best-fit of the ESO-H$\alpha$ 574 [\feii] lines gives the following parameters : $A_{\rm{V}}$ = 0 mag, $T_{\rm{e}}$ = 9\,000 K, and $n_{\rm{e}}$ = 2.0 10$^4$ cm$^{-3}$. 
A gas component at a single pair ($T_{\rm{e}}$, $n_{\rm{e}}$) fits reasonably well all the lines but systematically underestimates those coming from some doublets and sextets levels (b$^2$H, a$^6$S,  and a$^2$G), shown  with different colors in Figure\,\ref{fig:bestfit_eso} and reported in Table\,\ref{tab:tab3}. In particular, 
ratios involving lines from a$^6$S and a$^2$G levels (8 lines) are underestimated by a factor of two, while those from level b$^2$H (2 lines) are underestimated by a factor of four.
This systematic behavior, which can be reasonably ascribed to the poor knowledge of the atomic parameters, has been already evidenced by Bautista \& Pradhan (1998) for the a$^6$S level. Notably, however, the same model is selected as best-fit irrespective from including or not the doublets and sextets in the fit, although with a higher minimum reduced-$\chi^2$ (hereinafter $\chi^2$) in the latter case.\\
The sensitivity of the line ratios to the fitted parameters is probed in  Figure\,\ref{fig:chi_eso}, where we plot  the $\chi^2$-contours in the density-temperature plane for $A_{\rm{V}}$\,=\,0 mag (minimum $\chi^2$ = 0.9). 
Higher $A_{\rm{V}}$ values return fits with substantially higher $\chi^2$ and are therefore discarded (for example minimum $\chi^2$ = 1.9 for  $A_{\rm{V}}$\,=\,0.5 mag); this indicates that extinction decreases slightly from the ESO-H$\alpha$ 574 central source (where $A_{\rm{V}}$ $\sim$ 1.5 mag, WBA13) to the jet. The plotted contours refer to increasing $\chi^2$ values of 30\%, 60\%, and 90\% with respect to the minimum $\chi^2$ value.  
From this plot we derive that temperature and density do not exceed (inside a confidence of 3-\,$\sigma$) the ranges 8\,000 K $\la$ $T_{\rm{e}}$ $\la$ 11\,000 K and 6 10$^3$ cm$^{-3} \la $ $n_{\rm{e}}$ $\la $ 6 10$^4$ cm$^{-3}$, respectively. 

To check the reliability of the results, we have also attempted two different approaches: {\it i}) to fit the data with a different set of collisional coefficients (Bautista \& Pradhan 1998), which returns the same physical parameters but with a higher minimum $\chi^2$, and {\it ii}) to fit the ultra-violet component and the infrared components separately, with the aim to test the possibility of the presence of different gas components. 
The $\chi^2$-contours of the ultra-violet lines fit (Figure\,\ref{fig:chi_uv}) shows a best-fit value not significantly different from that obtained by the all-lines fit. Analogously, the temperature range does not significantly differ from that found in the all-lines fit. More interestingly, the density range traced by the ultra-violet lines points to higher densities (i.e. up to 10$^{5.8}$ cm$^{-3}$ within a 3-$\sigma$ confidence level). This suggests that while temperature is fairly constant in the probed region (or that its variations occur over spatial scales much smaller than the angular resolution), density may be subjected to stronger gradients. Finally, the fit of the infrared lines (not shown here) gives results in good agreement with the all-lines fit. 

In Figure\,\ref{fig:bestfit_parlup} we show the best-fit model for the [\feii] lines observed in Par-Lup 3-4. The minimum $\chi^2$ is found, as in the case of ESO-H$\alpha$ 574, by 
taking as a reference the 4277\,\AA\, line.
To better compare the line emission observed in this object with that of ESO-H$\alpha$ 574, we plot, together with the line ratios of the detected lines, also the 2-\,$\sigma$ upper limits at the wavelength of the lines detected only in  ESO-H$\alpha$ 574. As anticipated in Sec.\,\ref{sec:sec2}, in Par-Lup 3-4 the ratios between ultra-violet and optical/near-infrared lines are substantially higher.
This circumstance is a consequence of the higher temperature probed ($T_{\rm{e}}$\,=\,13\,000 K). The inferred electron density and extinction are $n_{\rm{e}}$ = 6.0 10$^4$ cm$^{-3}$ and $A_{\rm{V}}$\,=\,0 mag, respectively. As for ESO-H$\alpha$ 574, we find that the predictions of sextet and doublet levels are systematically underestimating  the observed ratios of a factor between two and three.  Within a confidence level of 90\%, the $\chi^2$-contour plot gives 11\,000 K $\la$ $T_{\rm{e}}$ $\la$ 20\,000 K and 1.8 10$^4$ cm$^{-3} \la $ $n_{\rm{e}}$ $\la $ 1.8 10$^5$ cm$^{-3}$ (see Figure\,\ref{fig:chi_par}).
Finally, if the collisional coefficients by Bautista \& Pradhan (1998) are adopted, the best-fit gives $T_{\rm{e}}$\,=\,16\,000 K, $n_{\rm{e}}$ = 8.0 10$^4$ cm$^{-3}$, $A_{\rm{V}}$= 0 mag.

\subsubsection{[\feiii] lines fit}{\label{sec:sec3.1.2}
The fit of [\feiii] lines detected in ESO-H$\alpha$ 574 is presented in Figure\,\ref{fig:fit_feiii}. The best-fit model is obtained by taking as a reference the line at 4930\,\AA\, (see also Table\,\ref{tab:tab2}). This gives the following parameters: $T_{\rm{e}}$ = 19\,000 K, $n_{\rm{e}}$= 2.0 10$^5$ cm$^{-3}$, $A_{\rm{V}}$ = 0 mag. At variance with Fe$^+$ lines, lines of Fe$^{++}$ lie all in the ultra-violet range and come all from levels with similar upper energy. Consequently, we expect that Fe$^{++}$ lines are poorly sensitive to the temperature. This is clear in the $\chi^2$-contour plot of Figure\,\ref{fig:chi_eso_feIII}, where all temperatures in the grid of NLTE solutions above 8 000\,K 
are compatible with the observations (within a confidence level of 90\%). Conversely, the electron density is better constrained within the range  1 10$^5$ cm$^{-3} \la $ $n_{\rm{e}}$ $\la $ 6 10$^5$ cm$^{-3}$. This result confirms that indeed a density gradient exists along the jet of ESO-H$\alpha$ 574, and that [\feiii] and [\feii] ultra-violet lines likely probe the same, high-density gas component.

\subsection{The ionization model}{\label{sec:sec3.2}

To consistently interpret the [\feii] and [\feiii] emission in ESO-H$\alpha$ 574 and to derive the fractional abundance Fe$^+$/Fe$^{++}$, 
we applied a ionization equilibrium code that involves the first 4 ionization 
stages of iron. The following processes have been taken into account: direct ionization, radiative and dielectronic recombination 
(data from Arnaud \& Raymond 1992), and direct and inverse charge-exchange with hydrogen (data from Kingdon \& Ferland 1996). Notably, while the first three processes are a function only of the electron temperature,  direct and inverse charge-exchange rates also depend on the  fractional ionization $x_{\rm{e}}$\,=\,$n_{\rm{e}}$/$n_{\rm{H}}$, where
 $n_{\rm{H}}$=$n_{\rm{H^0}}$+$n_{\rm{H}^+}$. Moreover, since the electron transfer is more efficient when the involved ions (e.g. H$^{0}$ and Fe$^+$) have similar ionization potentials (IP)\footnote{Being IP (Fe$^{0}$) = 7.87 eV,  IP (Fe$^{+}$) = 16.18 eV  and IP (Fe$^{++}$) = 30.64 eV, to be compared with IP(H$^{0}$) = 13.595 eV.}, the charge-exchange rate is relevant
only for the process Fe$^{+}$ + H$^{+}$ $\rightleftarrows$  Fe$^{++}$ + H$^{0}$. Therefore, it returns relevant {\bf results} for the  Fe$^{+}$/Fe$^{++}$ abundance ratio, while it is negligible for both the Fe$^{0}$/Fe$^{+}$ and Fe$^{++}$/Fe$^{+3}$ abundance ratios.\\
For $T_{\rm{e}}$= 8\,000 K, namely the lowest temperature derived from the $\chi^2-$contours of Figures\,\,\ref{fig:chi_eso} and \ref{fig:chi_uv}, our model predicts a substantial
fraction of iron in neutral form even if the gas is almost fully ionized (e.g. we get 30\% of Fe$^0$, 52\% of Fe$^+$, and 18\% of Fe$^{++}$ for $x_{\rm{e}}$ = 0.9). This strongly contrasts with the simultaneous lack of any Fe$^0$ line in the ESO-H$\alpha$ 574 spectrum together with the presence of Fe$^{++}$ lines. However, just a slight increase of the electron temperature at 9\,000 K makes the neutral Fe$^0$ percentage drop to less than 10\%, and that of Fe$^{++}$ to increase to more than 20\%, in agreement with the observations. The expected percentage of Fe$^{+3}$ is negligible for the whole range of temperature considered in Figures\,\ref{fig:chi_eso} and \ref{fig:chi_uv}.\\
To derive $x_{\rm{e}}$ we solved the ionization equilibrium equations (together with the excitation 
equilibrium for each of the two species) to predict a number of [\feii]/[\feiii] line ratios. We constructed a grid of model solutions in the range 0 $\le$ $x_{\rm{e}}$ $\le$ 1 (in
 steps of $\delta x_{\rm{e}}$\,=\,0.05) and  9\,000 K $\le T_{\rm{e}} \le$ 14\,000 K, being the upper value that derived from the $\chi^2-$contours of Figure\,\ref{fig:chi_uv}. To estimate  $x_{\rm{e}}$ we consider the
[\feii] ultra-violet lines and the [\feiii] lines, assuming that they come from the same portion of the post-shock gas (see Sect.\,\ref{sec:sec3.1.2}). 
We consider 14 line ratios involving 7 [\feii] lines with two bright [\feiii] lines  at 4701.59 \AA\, and 5270.53 \AA. As an example, we show in Figure\,\ref{fig:fit_xe}, upper panel, the [\feii]4244/[\feiii]5270 ratio as a function of $x_{\rm{e}}$ for the considered range of temperature. The observations are in agreement with  0.65 $\la$ $x_{\rm{e}}$ $\la$ 0.85, where the lower (upper) value refers to the highest (lowest) temperature assumed. This value of $x_{\rm{e}}$ is the same found (within the error range) if all the 14 ratios are considered.\\
For Par-Lup 3-4 we can derive an upper limit on x$_e$ by considering the upper limits on the [\feiii] lines. Taking a grid in the range 11\,000 $< T_{\rm{e}} <$ 20\,000 K (see Sect.\,\ref{sec:sec3.1.1} and Figure\,\ref{fig:chi_par}), we get $x_{\rm{e}}$  $\la$ 0.4. As an example, the derivation of $x_{\rm{e}}$ from the ratio [\feii]4244/[\feiii]5270) is shown in Figure\,\ref{fig:fit_xe}, lower panel.\\
Typical $x_{\rm{e}}$ values in protostellar jets range from 0.03 to 0.6 (Ray et al. 2007, Nisini et al. 2005, Podio et al. 2009), although x$_e$= 0.8 is found in the High Velocity Component (HVC) of the DG Tau B jet Podio et al (2011). Therefore, while the fractional ionization  of Par-Lup 3-4 is in the range of the most common values, that of ESO-H$\alpha$ 574 appears remarkably high.

\subsection{Photoexcitation contribution}{\label{sec:sec3.3}
In Section\,\ref{sec:sec3.1.1} the observed line ratios have been interpreted in the light of collisional excitation. In this Section we explore whether an additional contribution from fluorescence excitation 
can be relevant. In ESO-H$\alpha$ 574 this possibility is supported by the detection of bright [Ni\,II] lines at  7377.8 \AA\, and 7411.6 \AA \,(WBA13), whose
intensity is easily enhanced because of the pumping of an ultra-violet field (Lucy 1995), though the observed intensity ratio of around 10, is compatible only with collisional excitation (see Figure\,2 of Bautista et al. 1996). 
In Par-Lup 3-4, only the 7377.8 \AA\, line is detected.\\
To better investigate the role of photo-excitation in ESO-H$\alpha$ 574, we have included in the excitation model a radiation field, which can be produced either from the stellar photosphere or by a hot spot on the stellar surface produced by the accretion shock of the infalling matter.  
Both these fields have been  approximated as W\,$\times$\,{\it B$_{\nu}$ (T$_{\rm{eff}}$)}, where $B_{\nu}$ is the black-body function at the stellar (or hot spot) temperature and W=\,1/4\,(R/r)$^2$ is the dilution factor, having adopted the stellar radius {\rm R}= 3\,R$_{\odot}$ and the distance of the knot A1 from the star, {\rm r}, equal to 100 AU (i.e. 0.2$^{\prime\prime}$, see BWA11). We take T$_{\rm{eff}}$ = 4000 K for the stellar temperature and 6\,000 K$ \le$ T$_{\rm{eff}}$ $\le$ 12\,000 K for the hot spot temperature, following the model of Calvet \& Gullbring (1998). The hot spot area has been taken between 10-30\% of the stellar surface.\\
As shown by Lucy (1995), a powerful way to evaluate the relevance of photo-excitation, is to compute the so-called excitation parameter (U$_{\rm{ex}}$), which is defined as the ratio between all the radiative and collisional excitation rates involving two given levels. From U$_{\rm{ex}}$, the 'second critical electron density' can be also derived, $n^*_{\rm{e}}$ =  U$_{\rm{ex}}$ $n_{\rm{e}}$,  such that for $n^*_{\rm{e}}$ $\gg$  $n_{\rm{e}}$, fluorescent excitation is predominant with respect to collisional excitation. 
Assuming a stellar field and for $n_{\rm{e}}$ = 2\,10$^4$\,cm$^{-3}$ (Sect.\,\ref{sec:sec3.1.1}), we get $n^*_{\rm{e}}$ $\la$ 10$^2$ cm$^{-3}$ (or U$_{\rm{ex}}$ $<$ 5 10$^{-3}$) for all the levels, indicating that fluorescence excitation is negligible in this case. The importance of the
 hot-spot field was tested by varying both T$_{\rm{eff}}$ and W in the ranges given above, obtaining $n^*_{\rm{e}}$ up to 10$^5$ cm$^{-3}$. Thus, in principle, the presence of a hot-spot could have a role in fluorescence excitation. However, the comparison of the predicted intensity ratios with those observed in the ESO-H$\alpha$ 574 spectrum, indicates a marginal compatibility only for the lowest values of T$_{\rm{eff}}$ and W  (i.e. T$_{\rm{eff}}$ $\le$ 8\,000 K and hot-spot area not exceeding 10\% of the stellar surface). Hence, even if a hot-spot may exist, certainly it is not the main cause of the observed emission.
As a note, and with reference to Sect.\,\ref{sec:sec3.1.1} and Table\,\ref{tab:tab3}, we also
report that none  of the line ratios systematically underestimated by the collisional model can be reproduced even if fluorescence excitation is considered.\\
Finally, in the Par-Lup 3-4 case the distance between the central source and the jet is not well defined as in the ESO-H$\alpha$ 574 case. 
Taking different values of W, we estimate that photo-excitation contribution, and in particular that due to the hot spot field, can be relevant for distances closer than 5-10 AU from the central source. 
 
\section{Discussion}

\subsection{Comparison with shock models}

Once derived the physical conditions, the origin of the iron emission in the two jets was investigated in the framework of shock models. Figure\,\ref{fig:fit_hart}, adapted from Figure\,1 of Hartigan, Raymond, \& Morse (1994),  shows the variation of the ionization fraction, electron density, and temperature with the distance behind the shock front for a low velocity (35 km s$^{-1}$) and an intermediate velocity (70 km s$^{-1}$) shock, in the approximation of a slab geometry and for assumed values of the pre-shock density and magnetic field. For each combination of these parameters, we computed the intensity of the most prominent iron lines, then deriving their expected intensity variation along the overall post-shock region. In particular we show, in the left panels,
the peak-normalized intensity profiles of ultra-violet, optical, near-infrared [\feii] lines (those coming from levels a$^4$G, a$^4$P, and a$^4$D), and in the middle panels the profiles of [\feiii] lines coming from level a$^3$F. Notably, lines at different wavelengths peak at different distance from the shock front, in the dimensional scale of $\sim$ 10$^{13}$ - 10$^{14}$ cm. At the distance of our objects these scales correspond to hundredths of arcsec, which are not resolved at our spatial resolution, and therefore the excitation model of Fe$^{+}$ gives only average quantities. \\
It is also important to notice that the physical parameters derived in ESO-H$\alpha$ 574 and Par-Lup 3-4 cannot be directly compared with those depicted in Figure\,\ref{fig:fit_hart},
which strongly depend on the assumed conditions of pre-shock density of the gas, magnetic field strength and shock velocity. Nevertheless, a trend between post- and pre-
shock parameters can be evidenced. We computed (see Table\,\ref{tab:tab5}) the average $<T_{\rm{e}}>$, $<x_{\rm{e}}>$, $<n_{\rm{e}}>$ and the compression factor $C$ = $n_{\rm{post-shock}}/n_{\rm{pre-shock}}$, weighted by the intensity profiles of the various (groups of) lines depicted in Figure\,\ref{fig:fit_hart}. By examining  the data of Table\,\ref{tab:tab5}, a number 
of conclusions can be drawn: 1) for a given shock velocity, lines at decreasing wavelengths trace progressively higher temperatures. Ionization fraction and electron density slightly increase with decreasing wavelength in 
the model with $v_{\rm{shock}}$=70 km s$^{-1}$, while they remain fairly constant and significantly lower  if $v_{\rm{shock}}$=35 km s$^{-1}$; 2) the average parameters 
probed by the mean of all [\feii] lines (fourth line of Table\,\ref{tab:tab5}) indicate that increasing shock velocities correspond to decreasing temperatures and to increasing ionization fraction, electron density and compression factor. 
This points toward a higher shock-velocity in ESO-H$\alpha$ 574, 
where temperature is lower and electron density and ionization fraction are higher than in Par Lup 3-4 (see Table\,\ref{tab:tab4}). Moreover, in the intermediate-velocity shock model, the [\feiii] lines trace more specifically the portion of the post-shock region extending up to $\sim$ 10$^{13}$ cm behind the shock front, where the electron density reaches its maximum value. 
This region should therefore correspond to that traced by the observed [\feiii] line ratios.\\
We also note that the above scenario is also consistent with the abundance ratios of the Fe$^{0}$, Fe$^{+}$, and Fe$^{++}$ depicted in the right panels of Figure\,\ref{fig:fit_hart}.
Indeed, while for a low-velocity shock the bulk of iron is singly ionized, for an intermediate velocity shock the ratio Fe$^{+}$/Fe$^{++}$ $\sim$ 8 (at distances of the order of 
10$^{13}$ cm), again consistent with the detection of Fe$^{++}$ only in ESO-H$\alpha$ 574.\\
Finally, we again remark that although the above analysis allows us to interpret the observations in a consistent framework of shocked origin, the pre-shock parameters of the two models taken as a reference are not consistent with the derived post-shock parameters. For example, for the measured $<n_{\rm{e}}>$ and the compression factors of Table\,\ref{tab:tab5}, the pre-shock density would be 
$<n_{\rm{0}}>$ $\sim$ 7 10$^3$ cm$^{-3}$ and  $\sim$ 6 10$^4$ cm$^{-3}$ for ESO-H$\alpha$ 574 and Par Lup 3-4, respectively, which are higher than the $n_{\rm{0}}$ values at which the two models of Hartigan et al. (1994) are computed.

\subsection{Gas-phase Fe abundance}
The gas-phase Fe abundance $x$(Fe) is an indirect measure of the presence of dust inside the jet. In general jet launching models predict that the jet is dust-free as dust is completely destroyed in the launching region by the stellar radiation. Conversely, if the jet originates from a disk region  extending beyond the dust evaporation radius, it could
eventually transport some dust. This, in turn, could be then partially destroyed by the shock because of vaporisation and sputtering of energetic particles (e.g. Seab 1987; Jones 1999: Guillet et al. 2009). The degree of iron depletion is therefore also a function of the shock efficiency.
Previous studies of  $x$(Fe) in shock environments have given sparse results, from values close to solar abundance (e.g. Beck-Winchatz et al.
1996), up to intermediate (Nisini et al. 2002, Podio et al. 2006, 2009) and very high depletion factors (Mouri \& Taniguchi 2000; Nisini et al. 2005).  A powerful way to estimate the percentage
of gas-phase iron  ($\delta_{Fe}$), relies on intensity ratios involving lines of non-refractory species emitted in similar excitation conditions, as for example the [\feii]1.25$\mu$m/[\pii]1.18$\mu$m,
as suggested by Oliva et al. (2001).
Since phosphorous lines are not detected in our spectra, we investigate the possibility of using ratios involving [\oi] lines. To this aim, we solved the equations of ionization equilibrium for the first three ionic stages of oxygen, together with the statistical equilibrium for the first five levels of O$^0$. The radiative coefficients are taken from the 
NIST database\footnote{available at http://www.nist.gov/pml/data/asd.cfm} while the rates for collisions with electrons are from Bhatia \& Kastner (1995). As a result, we get the percentage of neutral oxygen and the peak-normalized intensity profile along the post-shock region. In particular, that of  {\ion{[O}{1}]} 6300\,\AA\, shown in the middle panels of Figure\,\ref{fig:fit_hart}, well resembles that of [\feii] ultra-violet lines. Therefore, we conclude that [\feii] ultra-violet lines and  {\ion{[O}{1}]} 6300\,\AA\ trace the same shock region and are therefore suited to measure  $\delta_{Fe}$ inside the shock.  This is also roughly confirmed by the average parameters traced by the {\ion{[O}{1]} optical lines reported in Table\,\ref{tab:tab5} and taken from Bacciotti \& Eisl\"{o}ffel (1999). Note also that other tracers commonly used to derive  $\delta_{Fe}$, such as [\sii] 6740\AA\,, are not as powerful as {\ion{[O}{1}]} 6300\,\AA\, since their shock profile does not resemble that of any iron line (see e.g. Figure 3 of Bacciotti \&  Eisl\"{o}ffel, 1999). The same problem arises if the {\ion{[O}{1}]} 6300\,\AA\, is
compared with [\feii] near-infrared lines (see Figure\,\ref{fig:fit_hart}).\\
To derive $\delta_{Fe}$, we thus selected  several ratios  [\oi] 6300\AA\footnote{The flux of [\oi] 6300\AA \, is (116.0$\pm$0.2) 10$^{-17}$ erg s$^{-1}$  cm$^{-2}$ in  ESO-H$\alpha$ 574  and (248.3$\pm$0.3) 10$^{-17}$ erg s$^{-1}$  cm$^{-2}$ in Par-Lup 3-4.}  over bright ultra-violet [\feii] lines, whose observed values  are compared with those expected for the $<T_{\rm{e}}>$, $<n_{\rm{e}}>$ and $<x_{\rm{e}}>$ determinations derived from the iron analysis. By assuming the solar iron and oxygen abundances with respect to hydrogen of 3.16 10$^{-5}$ and 6.76 10$^{-4}$ (Grevesse \& Sauval 1998), we estimate $\delta_{Fe}$ = 0.55 $\pm$ 0.05  and $\delta_{Fe}$ = 0.30 $\pm$ 0.03 for ESO-H$\alpha$ 574 and Par-Lup 3-4, respectively. This result is in agreement with the shock interpretation given in the previous section. The higher efficiency in destroying the dust in the shock in ESO-H$\alpha$ 574 is due to its higher velocity, as expected from models of dissociative shocks (Guillet et al. 2009). In this respect, further observational evidence is provided by the detection in ESO-H$\alpha$ 574 of bright lines from other refractory species, such as Ca and Ni, which, on the contrary, are barely detected in Par-Lup 3-4 (BWA11, WBA13). Finally, we note that the derived values of $\delta_{Fe}$ belong to the group of 'intermediate' depletion values, where the shock has not a sufficient strength to completely destroy dust. The presence of dust inside the shock is in turn an indication that the jet launching region is larger than the dust sublimation zone.  

\subsection{Comparison with the diagnostics of other atomic species}
Together with iron lines, the spectra of ESO-H$\alpha$ 574 and Par-Lup 3-4 are rich {\bf in} other atomic emission lines (BWA11, WBA13), some of which commonly used to diagnose the physical conditions of the emitting gas. In this section we intend to compare the parameters derived from iron lines with those traced by ratios of lines of oxygen, nitrogen and sulphur. To derive the theoretical values of such ratios we have implemented simple NLTE codes for the lowest 5 fine structure levels of each species. The radiative coefficients are taken
from the NIST database, while the electronic collision coefficients are taken from Pradhan (1976, [\oii]),  Pequignot, \& Aldovrandi (1976, [\n]), Mendoza (1983, [\nii]), 
Hollenbach, \& McKee (1989, [\sii]). The main results of this analysis, which are summarized in Table\, \ref{tab:tab6} are the following: {\it i}) on average the temperature probed in ESO-H$\alpha$ 574 is in agreement with that probed with iron lines. In Par-Lup 3-4 the derived temperatures give sparse results, with $T_{\rm{e}}$([\oi]) lower than 
$T_{\rm{e}}$([\feii]) and with  $T_{\rm{e}}$([\sii]) not consistent with $T_{\rm{e}}$([\nii]); {\it ii}) ratios of different species probe different electron densities, with $n_{\rm e}$([\oii]) $>$  $n_{\rm e}$([\n]) $>$ $n_{\rm e}$([\sii]). This result can be explained by comparing the fitted values with the critical densities of the involved lines, which, at $T_{\rm{e}}$ = 10\,000 K are of $\sim$ 10$^8$ cm$^{-3}$, $\sim$ 10$^6$  cm$^{-3}$,
and  $\sim$ 10$^4$  cm$^{-3}$ for [\oii], [\n] and [\sii], lines, respectively. While the densities traced with the [\sii] ratio are close to the critical value, and therefore not completely reliable, this is not the case for the density indicated by the [\oii] flux ratio. In ESO-H$\alpha$ 574 this density is the same as that inferred from the [\feiii] and [\feii]
ultra-violet lines, thus again supporting the result of a density gradient inside the jet. Notably, the [\oii] line ratio indicates that in Par-Lup 3-4 the density is higher than in ESO-H$\alpha$ 574, in agreement with what found with the [\feii] VIS and NIR lines.\\
In conclusion, care should be taken to compare physical conditions derived from different atomic
species and lines, due to the their different sensitivity to variations of physical parameters
behind the shock front. In this respect, the rich iron spectrum from UV to NIR, with lines sensitive to a large range
of excitation conditions, is particularly suited to obtain a more complete view of the post-shock
cooling region. 


%
\section{Summary}
We have analyzed the 3\,000-25\,000 \AA \,, X-shooter spectra, of two jets driven by low-luminosity pre-main sequence stars, ESO-H$\alpha$ 574 and Par-Lup 3-4, with
the aim of investigating the diagnostic capabilities of the iron lines. Our analysis and main results can be summarized as follows:
\begin{itemize}

\item[-] The spectra of the two objects are both rich in iron emission. More than 70 lines are detected in ESO-H$\alpha$ 574, (knot A1, up to 2$^{\prime\prime}$ from the source), while around 35 lines
are detected in the Par-Lup 3-4 jet (integrated up to 1$^{\prime\prime}$ from the source). The spectra show substantially different features. While in the 
Par-Lup 3-4 jet only [\feii] lines are detected, the spectrum of ESO-H$\alpha$ 574  shows both [\feii] and [\feiii] emission. The [\feii] lines are detected over the whole spectral
range, coming from levels with energy up to more than 30\,000 cm$^{-1}$. While in ESO-H$\alpha$ 574 the low-excitation, near-infrared lines are stronger than the
high-excitation, ultra-violet lines, the opposite occurs in Par-Lup 3-4.

\item[-] Both [\feii] and [\feiii] line ratios are interpreted through NLTE models. These allow {\bf us} to derive both the gas parameters (electron density and temperature)
along with the visual extinction. The [\feii] line fit indicates that the jet driven by ESO-H$\alpha$ 574 is, on average, colder ($T_{\rm e}$ $\sim$ 9\,000 K) and
less dense ($n_{\rm e}$ $\sim$ 2 10$^4$ cm$^{-3}$) than the Par-Lup 3-4 jet ($T_{\rm e}$ $\sim$ 13\,000 K, $n_{\rm e}$ $\sim$ 6 10$^4$ cm$^{-3}$). A more compact 
component ($n_{\rm e}$ $\sim$ 2 10$^5$ cm$^{-3}$) inside the jet is revealed in ESO-H$\alpha$ 574 if the ultra-violet lines are fitted separately from the optical and near-infrared lines. This
component, whose temperature is not well constrained, is likely the same responsible for the [\feiii] line emission.
The extinction appears to be negligible in both jets.

\item[-] The contribution of fluorescence excitation due to photons emitted from the central star was investigated. In ESO-H$\alpha$ 574 this effect is negligible, while it can have a role in Par-Lup 3-4 up to distances less than 10 AU from the central star.

\item[-] A ionization equilibrium code was applied to derive the fractional ionization ($x_{\rm e}$) inside the two jets. We get $x_{\rm e}$ $\sim$ 0.7 in ESO-H$\alpha$ 574 
and $x_{\rm e}$ $\la$ 0.4 in Par-Lup 3-4. In particular the value detected in ESO-H$\alpha$ 574 is remarkably high, as expected in high-velocity shocks. 

\item[-] The observational differences evidenced in the iron spectra of the two jets have been qualitatively interpreted in the framework of shock models. The physical parameters derived from the excitation analysis are consistent with shocks with different velocities, with the shock of ESO-H$\alpha$ 574 being significantly faster than that of Par-Lup 3-4. Plots of post-shock [\feii] line intensities vs. distance from the shock front indicate that lines at different wavelengths trace different post-shock regions.  In particular   
[\feii] ultra-violet and [\feiii] lines are emitted only close to the shock front (within a distance of $\sim$ 10$^{13}$ cm), where the post-shock density reaches its maximum value.

\item[-] The shock strength of the jets is probed by measuring the gas-phase iron abundance ($\delta_{Fe}$). This was derived from the ratios of fluxes of ultra-violet [\feii] lines with that of [\oi]\,6\,300\,\AA\,. Under the assumption of solar Fe and O abundances, we derive $\delta_{Fe}$ $\sim$
0.55 and 0.30 in  ESO-H$\alpha$ 574 and Par-Lup, respectively. This evidence is in agreement with the higher shock-velocity of ESO-H$\alpha$ 574, which in turn corresponds in a higher kinetic energy able to partially destroy the dust particles.

\item[-] The gas diagnostic derived from iron lines was compared with that obtained from bright lines of other atomic species detected in the X-shooter spectra. Although the average trend of temperature and density is the same (with ESO-H$\alpha$ 574 colder than Par-Lup 3-4), the derived values are in general not consistent with each-other. We ascribe this
behavior to the low number of the used lines, able to cover a limited parameter range that depends on the specific line excitation energies and critical densities. Conversely, thanks both to the very rich spectrum of iron and to the wide spectral range covered with X-shooter, the analysis of iron lines allows us to get a very comprehensive and consistent view of the gas physics in the post-shock region. 
\end{itemize}
\section{Acknowledgments} We are grateful to Manuel Bautista and to an anonymous referee for their suggestions and constructive discussions. TG and JMA thank also G. Attusino.   
The ESO staff is acknowledged for support with the observations and the X-shooter pipeline.
%

\begin{deluxetable}{cccccc}
\tabletypesize{\footnotesize} 
\tablecaption{\label{tab:tab1} [\feii] lines }  
\tablewidth{0pt}
\tablehead
{Line id.                                     & $\lambda_{air}$   & E$_{up}$             & $(R\pm\Delta R)_{ESO-H\alpha574}^{*}$   & $(R\pm\Delta R)_{Par-Lup3-4}^{*}$ & $R_{E}/R_{P}^{*}$ \\
                                             &  ($\AA$)          &(cm$^{-1})$           &                                 &                                 &         }
\startdata
$\mathbf{b^4\!D_{5/2}}-$a$^4\!$D$_{5/2}$     &  4347.35          &  31387.9             &      0.4 $\pm$ 0.2$^{**}$       &  $<$ 0.3                        & -       \\
         b$^4\!D_{1/2}-$a$^4\!$D$_{1/2}$     &  4438.91          &  31368.4             &      0.4 $\pm$ 0.2$^{**}$       &  $<$ 0.3                        & -       \\
$\mathbf{a^2\!F_{7/2}}-$a$^4\!$D$_{7/2}^a$   &  5163.95          &  27314.9             &      0.8 $\pm$ 0.3              &  $<$ 0.3                        & -       \\
$\mathbf{b^2\!H_{11/2}}-$a$^4\!$F$_{9/2}$    &  4114.46          &  26170.2             &      1.1 $\pm$ 0.4              &  0.6 $\pm$ 0.3$^{**}$           & 2.7     \\
         b$^2\!H_{11/2}-$a$^4\!$F$_{7/2}$    &  4211.09          &  26170.2             &      0.5 $\pm$ 0.2              &  $<$ 0.7                        & -       \\
$\mathbf{a^4\!G_{7/2}}-$a$^4\!$F$_{5/2}$     &  4319.61          &  25981.6             &      0.7 $\pm$ 0.3              &  0.7 $\pm$ 0.2                  & 1.0     \\
         a$^4\!G_{9/2}-$a$^4\!$F$_{5/2}$     &  4352.77          &  25805.3             &      0.4 $\pm$ 0.2              &  0.5 $\pm$ 0.2                  & 0.8     \\
        a$^4\!G_{11/2}-$a$^4\!$F$_{9/2}$     &  4243.96          &  25428.8             &      2.4 $\pm$ 0.7              &  1.8 $\pm$ 0.4                  & 1.3     \\
        a$^4\!G_{11/2}-$a$^4\!$F$_{7/2}$     &  4346.85          &  25428.8             &      0.3 $\pm$ 0.2$^{**}$       &  0.6 $\pm$ 0.2                  & 0.5     \\
$\mathbf{a^6\!S_{5/2}}-$a$^6\!$D$_{9/2}$     &  4287.39          &  23317.6             &      2.6 $\pm$ 0.7              &  $<$ 0.4                        & -       \\
         a$^6\!S_{5/2}-$a$^6\!$D$_{7/2}$     &  4359.33          &  23317.6             &      1.9 $\pm$ 0.5              &  1.5 $\pm$ 0.3                  & 1.3     \\
         a$^6\!S_{5/2}-$a$^6\!$D$_{5/2}$     &  4413.78          &  23317.6             &      1.2 $\pm$ 0.4              &  1.2 $\pm$ 0.2                  & 1.0     \\
         a$^6\!S_{5/2}-$a$^6\!$D$_{3/2}$     &  4452.09          &  23317.6             &      1.1 $\pm$ 0.4              &  0.5 $\pm$ 0.2                  & 2.2     \\
 $\mathbf{b^4\!F_{3/2}}-$a$^4\!$F$_{5/2}$    &  4950.74          &  23031.3             &      0.4 $\pm$ 0.2              &  $<$ 0.2                        & -       \\
         b$^4\!F_{5/2}-$a$^4\!$F$_{5/2}$     &  4973.38          &  22939.4             &      0.4 $\pm$ 0.2              &  $<$ 0.3                        & -       \\
         b$^4\!F_{5/2}-$a$^6\!$D$_{7/2}$     &  4432.44          &  22939.4             &      0.5 $\pm$ 0.2              &  $<$ 0.3                        & -       \\
         b$^4\!F_{7/2}-$a$^6\!$D$_{7/2}$     &  4457.94          &  22810.4             &      0.9 $\pm$ 0.3              &  0.8 $\pm$ 0.2                  & 1.1     \\
         b$^4\!F_{7/2}-$a$^4\!$F$_{9/2}$     &  4774.71          &  22810.4             &      0.4 $\pm$ 0.2              &  $<$ 0.2                        & -       \\
         b$^4\!F_{7/2}-$a$^4\!$F$_{7/2}$     &  4905.33          &  22810.4             &      0.8 $\pm$ 0.3              &  0.7 $\pm$ 0.2                  & 1.1     \\
         b$^4\!F_{9/2}-$a$^6\!$D$_{9/2}$     &  4416.26          &  22637.2             &      1.8 $\pm$ 0.4              &  1.7 $\pm$ 0.3                  & 1.0     \\
         b$^4\!F_{9/2}-$a$^4\!$F$_{9/2}$     &  4814.53          &  22637.2             &      1.5 $\pm$ 0.4              &  1.1 $\pm$ 0.2                  & 1.4     \\
 $\mathbf{a^4\!H_{7/2}}-$a$^4\!$F$_{3/2}$    &  5376.45          &  21711.9             &      1.2 $\pm$ 0.3              &  1.3 $\pm$ 0.3                  & 0.9     \\
         a$^4\!H_{9/2}-$a$^4\!$F$_{5/2}$     &  5220.05          &  21581.6             &      0.8 $\pm$ 0.3              &  0.8 $\pm$ 0.2                  & 1.0     \\
         a$^4\!H_{9/2}-$a$^4\!$F$_{9/2}$     &  5333.64          &  21581.6             &      1.4 $\pm$ 0.4              &  1.6 $\pm$ 0.3                  & 0.9     \\
         a$^4\!H_{11/2}-$a$^4\!$F$_{9/2}$    &  5111.62          &  21430.4             &      0.8 $\pm$ 0.3              &  1.0 $\pm$ 0.2                  & 0.8     \\
         a$^4\!H_{11/2}-$a$^4\!$F$_{7/2}$    &  5261.62          &  21430.4             &      2.8 $\pm$ 0.7              &  2.1 $\pm$ 0.3                  & 1.3     \\
         a$^4\!H_{13/2}-$a$^4\!$F$_{9/2}$    &  5158.77          &  21251.6             &      6.3 $\pm$ 1.4              &  3.7 $\pm$ 0.6                  & 1.7     \\
$\mathbf{b^4\!P_{3/2}}-$a$^6\!$D$_{5/2}$     &  4728.06          &  21812.1             &      0.4 $\pm$ 0.2              &  0.7 $\pm$ 0.2                  & 0.6     \\
         b$^4\!P_{3/2}-$a$^6\!$D$_{1/2}$     &  4798.27          &  21812.1             &      0.3 $\pm$ 0.2$^{**}$       &  $<$ 0.2                        & -       \\
         b$^4\!P_{5/2}-$a$^6\!$D$_{7/2}$     &  4889.61          &  20830.6             &      1.5 $\pm$ 0.4              &  0.9 $\pm$ 0.2                  & 1.7     \\
         b$^4\!P_{5/2}-$a$^4\!$F$_{9/2}$     &  5273.34          &  20830.6             &      1.2 $\pm$ 0.3              &  0.9 $\pm$ 0.2                  & 1.3     \\  
$\mathbf{a^2\!D_{3/2}}-$a$^4\!$F$_{5/2}^b$   &  5412.65          &  21308.0             &      0.4 $\pm$ 0.2              &  0.4 $\pm$ 0.1                  & 1.0     \\             
         a$^2\!D_{5/2}-$a$^4\!$F$_{7/2}^c$   &  5527.33          &  20517.0             &      1.7 $\pm$ 0.5              &  1.9 $\pm$ 0.3                  & 0.9     \\
$\mathbf{a^2\!H_{11/2}}-$a$^4\!$F$_{9/2}$    &  5413.34          &  20340.3             &      0.5 $\pm$ 0.2              &  $<$ 0.3                        & -       \\


$\mathbf{a^2\!G_{7/2}}-$a$^4\!$F$_{7/2}$     &  7172.00          &  16369.4              &   2.9 $\pm$ 0.7                 & 1.6 $\pm$ 0.3                  & 1.8     \\      
         a$^2\!G_{7/2}-$a$^4\!$F$_{5/2}$     &  7388.17          &  16369.4              &   1.7 $\pm$ 0.4                 & 1.7 $\pm$ 0.3                  & 1.0     \\
         a$^2\!G_{9/2}-$a$^4\!$F$_{9/2}$     &  7155.16          &  15844.6              &  10.7 $\pm$ 2.2                 & 5.4 $\pm$ 0.8                  & 2.0     \\
         a$^2\!G_{9/2}-$a$^4\!$F$_{7/2}$     &  7452.54          &  15844.6              &   3.2 $\pm$ 0.7                 & 1.9 $\pm$ 0.3                  & 1.7     \\
$\mathbf{a^4\!P_{1/2}}-$a$^4\!$F$_{5/2}$     &  9033.49          &  13904.8              &   1.7 $\pm$ 0.5                 & $<$ 1.3                        & -       \\
         a$^4\!P_{1/2}-$a$^4\!$F$_{3/2}$     &  9267.56          &  13904.8              &   2.2 $\pm$ 0.5                 & $<$ 1.3                        & -       \\
         a$^4\!P_{5/2}-$a$^6\!$D$_{5/2}$     &  7637.50          &  13474.4              &   1.7 $\pm$ 0.4                 & 0.8 $\pm$ 0.3                  & 2.1     \\
         a$^4\!P_{5/2}-$a$^4\!$F$_{9/2}$     &  8616.95          &  13474.4              &  11.0 $\pm$ 2.2                 & 3.8 $\pm$ 0.7                  & 2.9     \\
         a$^4\!P_{5/2}-$a$^4\!$F$_{7/2}$     &  9051.94          &  13474.4              &   3.1 $\pm$ 0.7                 & $<$ 1.3                        & -       \\ 
         a$^4\!P_{3/2}-$a$^6\!$D$_{5/2}$     &  7686.93          &  13673.2              &   0.9 $\pm$ 0.3                 & $<$ 0.3                        & -       \\
         a$^4\!P_{3/2}-$a$^6\!$D$_{7/2}$     &  8891.91          &  13673.2              &   4.2 $\pm$ 0.9                 & 1.7 $\pm$ 0.7                  & 2.5     \\
         a$^4\!P_{3/2}-$a$^6\!$F$_{5/2}^d$   &  9226.61          &  13673.2              &   3.0 $\pm$ 0.7                 & 11.9 $\pm$ 2.0                 & -       \\   
$\mathbf{a^4\!D_ {1/2}}-$a$^4\!$F$_{5/2}$    &  16637.6          &   8846.8              &   2.9 $\pm$ 0.8                 & $<$ 2.3                        & -       \\
         a$^4\!D_{3/2}-$a$^4\!$F$_{3/2}$     &  12787.7          &   8680.4              &   4.5 $\pm$ 1.5                 & 2.1 $\pm$ 0.7                  & 2.1     \\ 
         a$^4\!D_{3/2}-$a$^4\!$F$_{1/2}$     &  12977.7          &   8680.4              &   1.5 $\pm$ 0.6                 & $<$ 3.1            	     	& -       \\
         a$^4\!D_{3/2}-$a$^4\!$F$_{7/2}$     &  15994.7          &   8680.4              &   5.7 $\pm$ 1.3                 & $<$ 1.7                        & -       \\
         a$^4\!D_{3/2}-$a$^4\!$F$_{3/2}$     &  17971.0          &   8680.4              &   1.5 $\pm$ 0.7                 & $<$ 1.7                        & -       \\
         a$^4\!D_{5/2}-$a$^6\!$D$_{7/2}$     &  12485.4          &   8391.9              &   0.9 $\pm$ 0.3                 & $<$ 3.1                        & -       \\
         a$^4\!D_{5/2}-$a$^6\!$D$_{5/2}$     &  12942.6          &   8391.9              &   5.5 $\pm$ 1.4                 & 1.9 $\pm$ 0.7                  & 2.9     \\
         a$^4\!D_{5/2}-$a$^6\!$D$_{3/2}$     &  13277.7          &   8391.9              &   3.8 $\pm$ 1.0                 & $<$ 3.1                        & -       \\        
         a$^4\!D_{5/2}-$a$^6\!$D$_{9/2}$     &  15334.7          &   8391.9              &   5.5 $\pm$ 1.3                 & $<$ 1.7                        & -       \\
         a$^4\!D_{5/2}-$a$^6\!$D$_{7/2}$     &  16768.7          &   8391.9              &   6.6 $\pm$ 1.5                 & $<$ 2.3                        & -       \\
         a$^4\!D_{7/2}-$a$^6\!$D$_{9/2}$     &  12566.8          &   7955.3              &  32.9 $\pm$ 6.8                 & 7.4$\pm$1.3                    & 4.4     \\
         a$^4\!D_{7/2}-$a$^6\!$D$_{7/2}$     &  13205.5          &   7955.3              &  10.7 $\pm$ 2.4                 & $<$ 3.1                        & -       \\
         a$^4\!D_{7/2}-$a$^4\!$F$_{9/2}$     &  16435.4          &   7955.3              &  28.9 $\pm$ 6.0                 & 6.3 $\pm$ 1.2                  & 4.6     \\
         a$^4\!D_{7/2}-$a$^4\!$F$_{7/2}$     &  18093.9          &   7955.3              &   6.3 $\pm$ 2.0                 & $<$ 3.1                        & -       \\
\enddata
\tablenotetext{*}{~ $R_{ESO-H\alpha574}$, $R_{Par-Lup3-4}$ are both computed with respect to the line a$^4\!G_{9/2}-$a$^4\!$F$_{7/2}$ at 4276.82 \AA, whose flux is (2.5$\pm$0.5) $\times$ 10$^{-17}$ erg s$^{-1}$ cm$^{-2}$ in ESO-H$\alpha$ 574, knot A1, and (1.7$\pm$0.2) $\times$ 10$^{-17}$ erg s$^{-1}$ cm$^{-2}$ in Par-Lup 3-4. $R_{E}/R_{P}$ is the ratio $R_{ESO-H\alpha574}/R_{Par-Lup3-4}$}
\tablenotetext{**}{~ Line with signal-to-noise ratio between 2 and 3.}
\tablenotetext{a}{~ Blended with [\crii] c$^2\!$F$_{7/2}$-a$^4\!$G$_{9/2}$ at 5164.45 \AA.}
\tablenotetext{b}{~ Blended with [\feiii] $^3\!$P$_{2}$-$^5\!$D$_{1}$ at 5412.08 \AA.}
\tablenotetext{c}{~ Blended with [\feii] b$^2\!$P$_{1/2}$-a$^4\!$D$_{1/2}$ at  5527.60 \AA.}
\tablenotetext{d}{~ In Par-Lup 3-4 blended with He I $^3\!$P$_{0}$-$^3\!$D$_{0}$.}
\end{deluxetable}


\begin{deluxetable}{cccc}
\tabletypesize{\footnotesize} 
\tablecaption{\label{tab:tab2} [\feiii] lines in ESO-H$\alpha$ 574- knot A1. }  
\tablewidth{0pt}
\tablehead
{Line id.                                     & $\lambda_{air}$   & E$_{up}$     & R$\pm\Delta^{*}R$       \\
                                             & ($\AA$)            &(cm$^{-1})$  &                             }
\startdata
   $\mathbf{^3\!F_{2}}-^5\!$D$_{2}$          &  4733.91          & 21857.2     &  0.8 $\pm$ 0.3             \\
         ${^3\!F_{2}}-^5\!$D$_{1}$           &  4777.68          & 21857.2     &  1.2 $\pm$ 0.5             \\
         $^3\!$F$_{3}-^5\!$D$_{4}$           &  4607.03          & 21699.9     &  1.3 $\pm$ 0.6             \\
         $^3\!$F$_{3}-^5\!$D$_{3}$           &  4701.53          & 21699.9     &  2.8 $\pm$ 0.7             \\
         $^3\!$F$_{3}-^5\!$D$_{2}$           &  4769.43          & 21699.9     &  1.4 $\pm$ 0.5             \\
         $^3\!$F$_{4}-^5\!$D$_{4}$           &  4658.05          & 21462.2     &  7.6 $\pm$ 1.6             \\
         $^3\!$F$_{4}-^5\!$D$_{3}$           &  4754.69          & 21462.2     &  1.8 $\pm$ 0.6             \\
$\mathbf{^3\!H_{4}}-^5\!$D$_{4}$             &  4881.00          & 20481.9     &  3.8 $\pm$ 1.0             \\
         $^3\!$H$_{4}-^5\!$D$_{3}$           &  4987.2           & 20481.9     &  1.1 $\pm$ 0.4             \\
  $\mathbf{^3\!P_{1}}-^5\!$D$_{2}$           &  5011.25          & 20688.4     &  1.8 $\pm$ 0.6             \\
         $^3\!$P$_{2}-^5\!$D$_{3}$           &  5270.40          & 19404.8     &  3.3 $\pm$ 0.8             \\
         $^3\!$P$_{2}-^5\!$D$_{1}^a$         &  5411.98          & 19404.8     &  1.2 $\pm$ 0.6             \\
\enddata       
\tablenotetext{*}{~ Te flux ratio R is computed with respect to the $^3\!F_{2}-^5\!$D$_{2}$ line at  4930.53 \AA, whose flux is (1.2$\pm$0.2) $\times$ 10$^{-17}$ erg s$^{-1}$ cm$^{-2}$.}
\tablenotetext{a}{~ Blended with [\feii] a$^2\!$D$_{3/2}$-a$^4\!$F$_{5/2}$ at 5412.65 \AA.}
\end{deluxetable}

\begin{deluxetable}{cc}
\tabletypesize{\footnotesize} 
\tablecaption{\label{tab:tab3} [\feii] lines coming from doublets and sextets levels underestimated by the NLTE model. }  
\tablewidth{0pt}
\tablehead
{Line id.                           & $\lambda_{air}$ (\AA)}
\startdata
b$^2\!H_{11/2}-$a$^4\!$F$_{9/2}$    &  4114.46      \\
b$^2\!H_{11/2}-$a$^4\!$F$_{7/2}$    &  4211.09      \\
a$^6\!S_{5/2}-$a$^6\!$D$_{9/2}$     &  4287.39      \\
a$^6\!S_{5/2}-$a$^6\!$D$_{7/2}$     &  4359.33      \\
a$^6\!S_{5/2}-$a$^6\!$D$_{5/2}$     &  4413.78      \\
a$^6\!S_{5/2}-$a$^6\!$D$_{3/2}$     &  4452.09      \\
a$^2\!G_{7/2}-$a$^4\!$F$_{7/2}$     &  7172.00      \\      
a$^2\!G_{7/2}-$a$^4\!$F$_{5/2}$     &  7388.17      \\
a$^2\!G_{9/2}-$a$^4\!$F$_{9/2}$     &  7155.16      \\
a$^2\!G_{9/2}-$a$^4\!$F$_{7/2}$     &  7452.54      \\
\enddata
\end{deluxetable}

\begin{deluxetable}{ccc}
\tabletypesize{\footnotesize} 
\tablecaption{\label{tab:tab4} Fitted physical parameters. } 
\tablewidth{0pt}
\tablehead
{                                        &  ESO-H$\alpha$ 574    & Par-Lup 3-4      }
\startdata
Temperature (10$^4$ K)                  &  0.8 - 1.4             & 1.1- 2.0        \\
Electron density (10$^4$ cm$^{-3}$)     &  0.8 - 63.0$^a$            & 1.8 -17.7          \\
Ionization fraction                     &  0.65-0.85             & $<$ 0.4       \\
Gas phase iron (\%)                     &  50-60                 &  27-33          \\
\enddata
\tablenotetext{a}{The upper value is derived from the fit of [{\feii}] ultra-violet lines.}
\end{deluxetable}

\begin{deluxetable}{lcccc|cccc}
\tabletypesize{\footnotesize} 
\tablecaption{\label{tab:tab5} Intensity-weighted parameters in the shock cooling region (computed from the models of Fig.\,\ref{fig:fit_hart}). }
\tablewidth{0pt}
\tablehead
{          &  \multicolumn{4}{c}{35 km s$^{-1}$}                           &  \multicolumn{4}{c}{70 km s$^{-1}$}  \\
Lines                      &  $<T_{\rm{e}}>$   &  $<x_{\rm{e}}>$    & $<n_{\rm{e}}>$    &  $<C>$ &  $<T_{\rm{e}}>$ &  $<x_{\rm{e}}>$ &  $<n_{\rm{e}}>$  & $<C>$   \\
                           &       (K)         &     -              &  (cm$^{-3}$)      &    -   &     (K)         &     -           &  (cm$^{-3}$)     &  -      }
\startdata
{\ion{[Fe}{2}]}   a$^4$D  &     8580          &   0.032            &  207              &  7     &    5690         &    0.26         &    3610          & 15      \\
{\ion{[Fe}{2}]}   a$^4$P  &     9000          &   0.033            &  207              &  6     &    7220         &    0.32         &    4380          & 14      \\
{\ion{[Fe}{2}]}   a$^4$G  &     9760          &   0.034            &  205              &  6     &    8870         &    0.36         &    4810          & 14      \\
{\ion{[Fe}{2}]}  all lines&     9110          &   0.033            &  206              &  6     &    7260         &    0.31         &    4270          & 14      \\     
{\ion{[Fe}{3}]}  a$^3$F  &    11900          &   0.034            &  197              &  6     &   14100         &    0.41         &    5020          & 13      \\ 
{\ion{[O}{1}]}$^a$         &     9090          &   0.033            &   -               &  -     &    9180         &    0.364        &     -            &  -      \\ 
\enddata
\tablenotetext{a}{~ Taken from Bacciotti \& Eisl\"{o}ffel, (1999).}
\end{deluxetable}

\begin{deluxetable}{c|cccc}
\tabletypesize{\footnotesize} 
\tablecaption{\label{tab:tab6} Diagnostics of other atomic lines } 
\tablewidth{0pt}
\tablehead
{    Ratio                              &    \multicolumn{2}{c}{ESO-H$\alpha$ 574}  &    \multicolumn{2}{c}{Par-Lup 3-4}    \\
                                        &     Obs. ratio         &  T$_e$ (K)       &    Obs. ratio             & T$_e$ (K)          }
\startdata                                      
{\ion{[O}{1}]}(6300+6363)/5577          &     26.0               &    11\,000       &    28.4                   &    9\,000         \\              
{\ion{[N}{2}]}(6548+6583)/5755          &     28.5               &    12\,000       &    $>$ 15                 &    $<$ 20\,000    \\
{\ion{[S}{2}]}(6716+6731)/(4076+4069)   &     2.8                &    12\,000       &    0.7                    &    $>$ 20\,000    \\
\hline
                                        &     Obs. ratio         & n$_e$ (cm$^{-3}$)&   Obs. ratio              & n$_e$ (cm$^{-3}$) \\
\hline
{\ion{[O}{2}]}(3726+3729)/(7319+7330)   &      1.4               &   2 10$^5$       &    0.4                    &  8 10$^5$         \\
{\ion{[N}{1}]}(5198+5200)/(10398+10407) &      2.4               &   1 10$^4$       &    0.1                    &  $>$ 10$^5$       \\
{\ion{[S}{2}]}6716/6731                 &      0.6               &   5 10$^3$       &    0.5                    &  10$^4$           \\
\enddata
\end{deluxetable}
%


\begin{figure*}
\includegraphics[width=15cm]{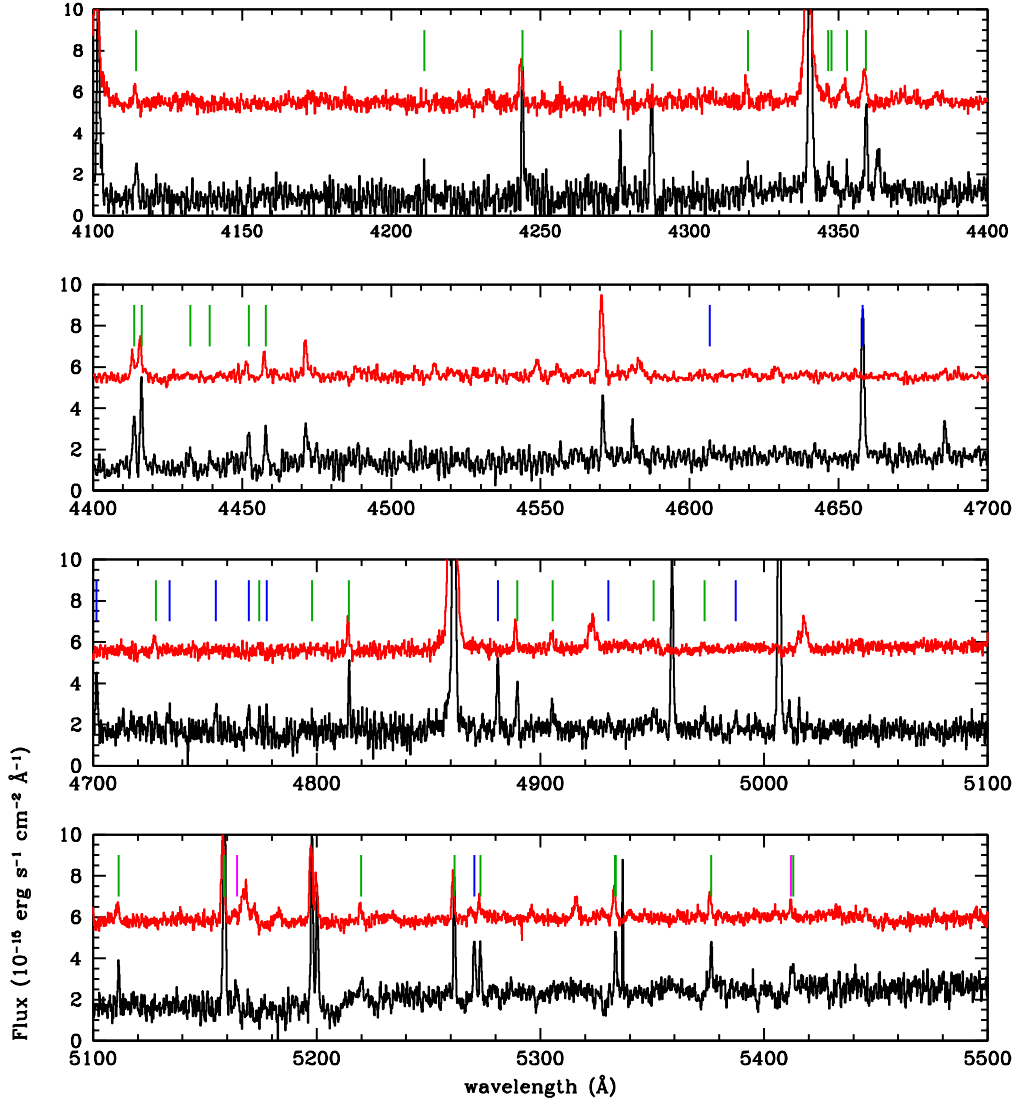}
\caption{\label{fig:spettro_uv} UVB spectrum of ESO-H$\alpha$ 574 (black) and Par-Lup 3.4 (red) where iron lines are detected. Green labels : [\feii] lines;
blue labels : [\feiii] lines: magenta labels: blends. For clarity, the spectrum of Par-Lup 3-4 was augmented by a factor of 5 (in the reported units).}
\end{figure*}
%

\begin{figure*}
\includegraphics[width=15cm]{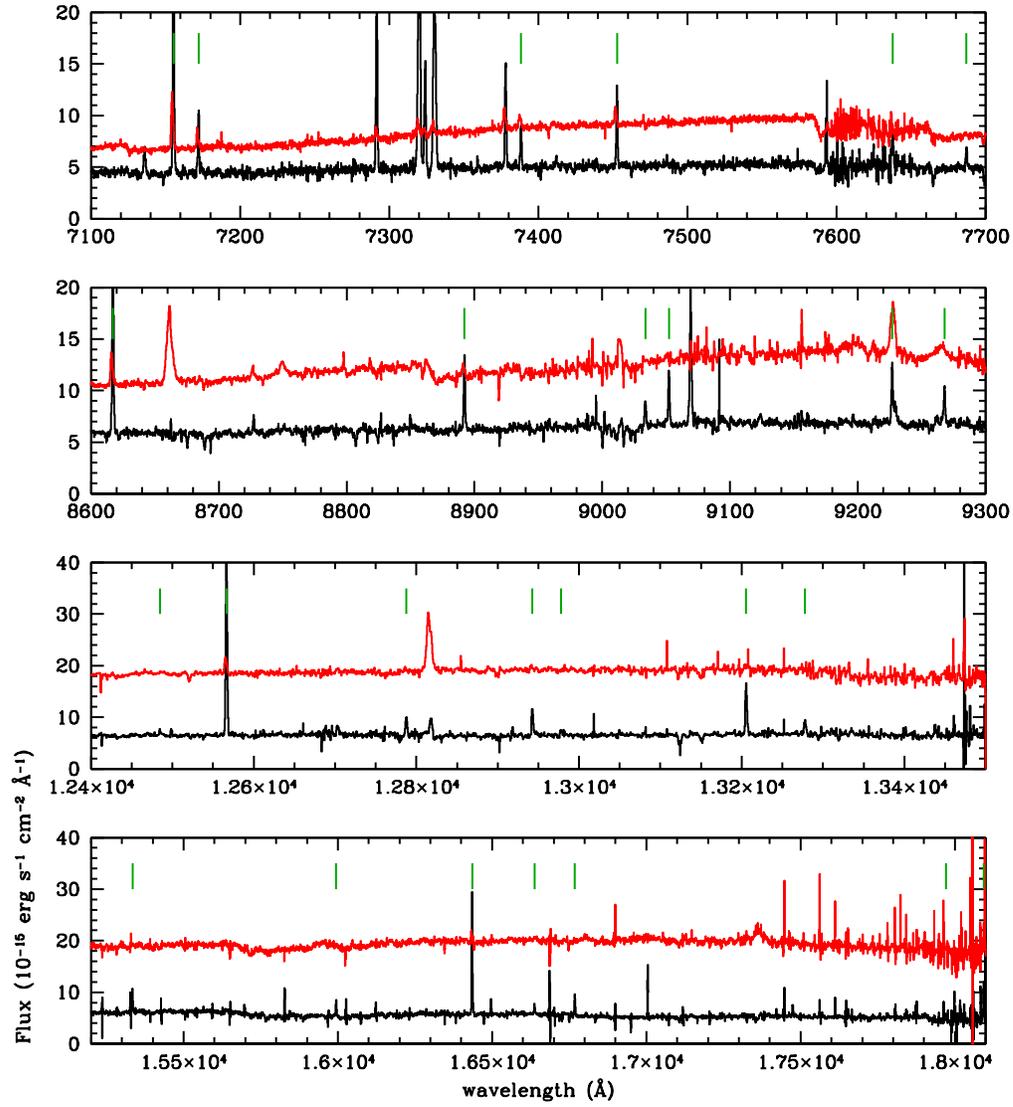}
\caption{\label{fig:spettro_visnir} As in Figure\,\ref{fig:spettro_uv} for the VIS and NIR spectra.}
\end{figure*}
%
\begin{figure*}
\includegraphics[width=15cm]{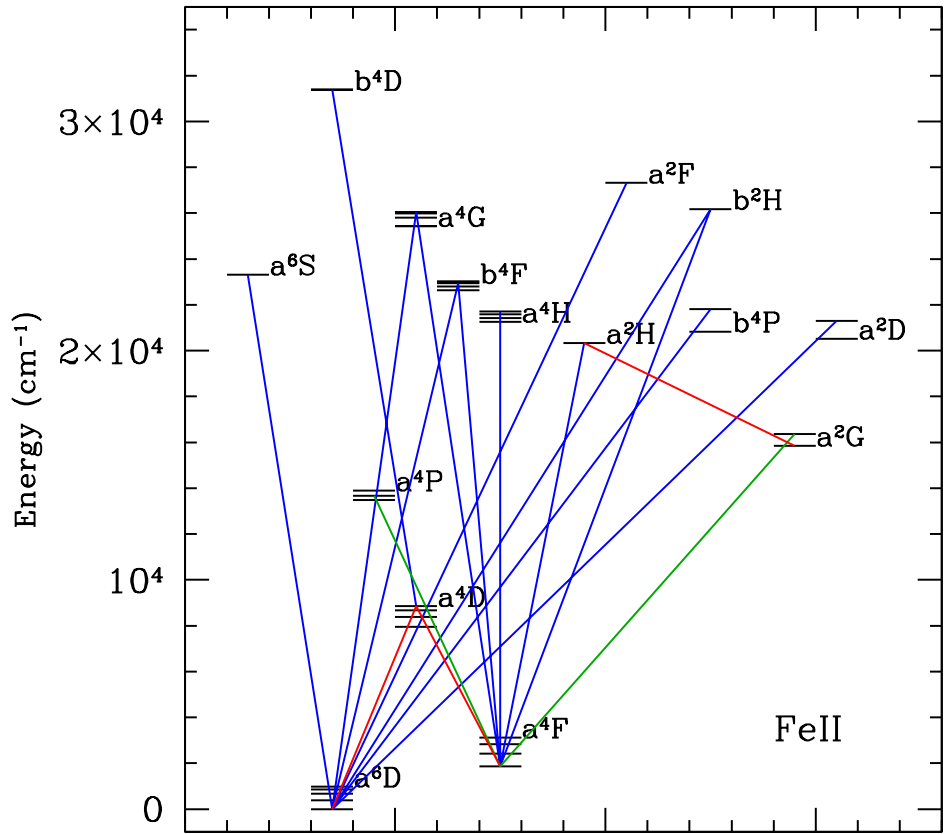}
\caption{\label{fig:livfeII} Grotrian diagram of Fe$^+$ levels associated with the observed lines.  Groups of lines detected in different X-shooter arms are depicted with different colors : blue: ultra-violet lines, green: optical
lines, red: near-infrared lines.}
\end{figure*}
%


\begin{figure*}
\includegraphics[width=15cm]{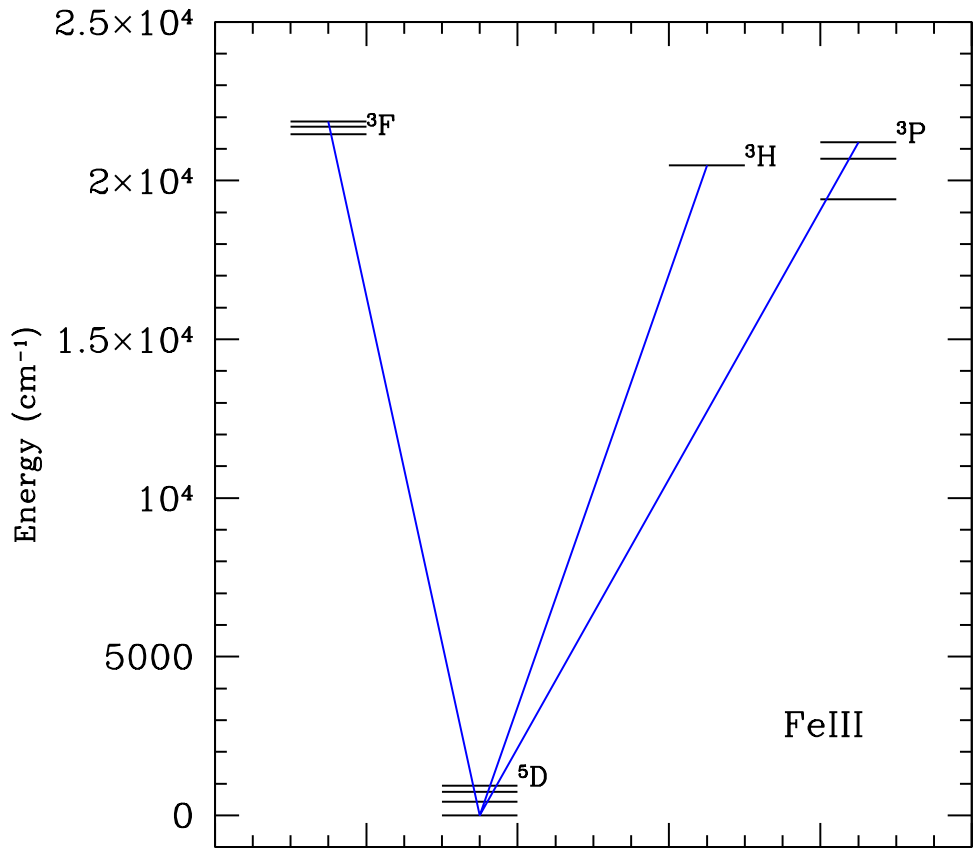}
\caption{\label{fig:livfeIII} As in Figure\,\ref{fig:livfeII} for the  Fe$^{++}$ levels.}
\end{figure*}

\begin{figure}
\includegraphics[width=15cm]{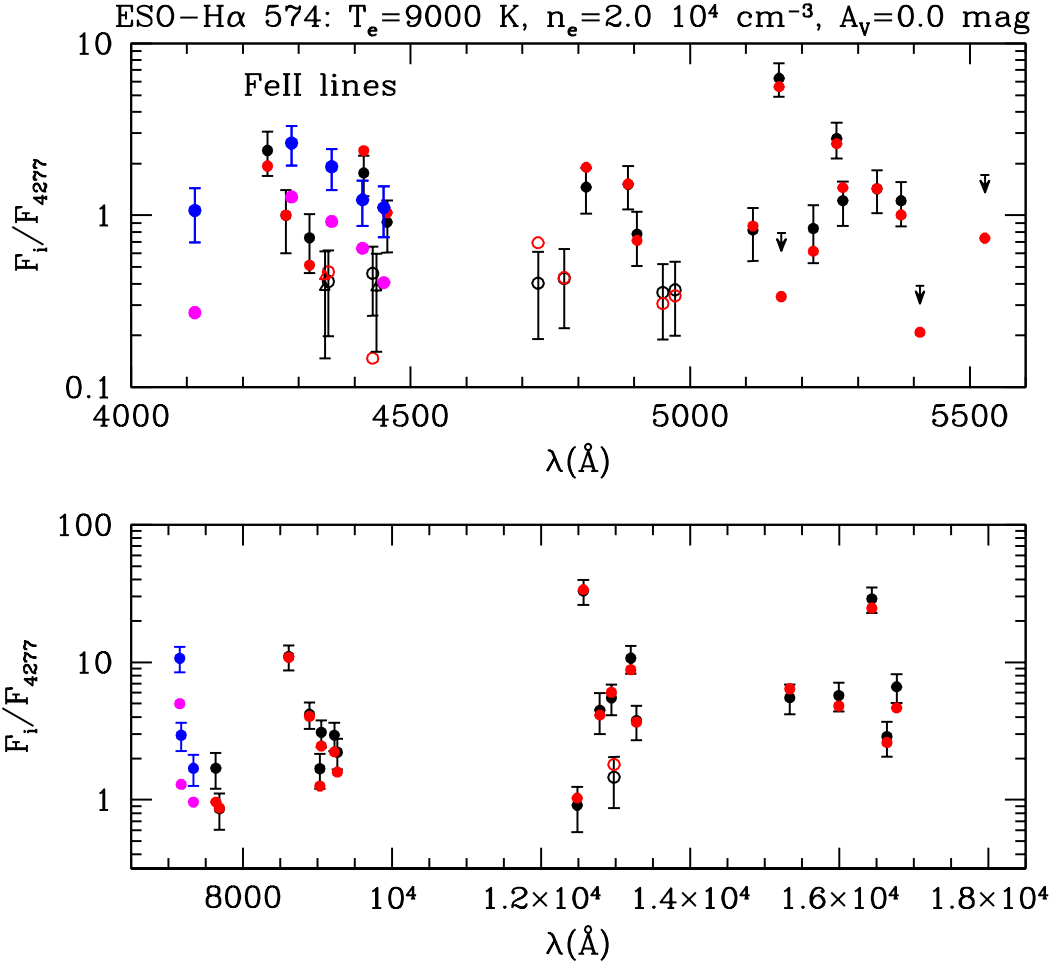}
\caption{\label{fig:bestfit_eso} NLTE best-fit model of the [\feii] lines  detected in ESO-H$\alpha$ 574. In the fitting procedure we have included the lines detected with snr $\ge$ 5, represented as filled circles (black: data, red: model). Lines detected with 3\,$\le$ snr $<$\,5 and 2\,$\le$ snr $<$\,3 are reported with open circles and open triangles, respectively.
Down arrows are the fluxes of blended lines. Blue filled circles and magenta filled circles indicate the observed data and model predictions of lines coming from levels b$^2$H, a$^6$S, and a$^2$G, which are not included in the fitting procedure (see text). The best-fit parameters are reported as well.}
\end{figure}

\begin{figure}
\includegraphics[width=10cm]{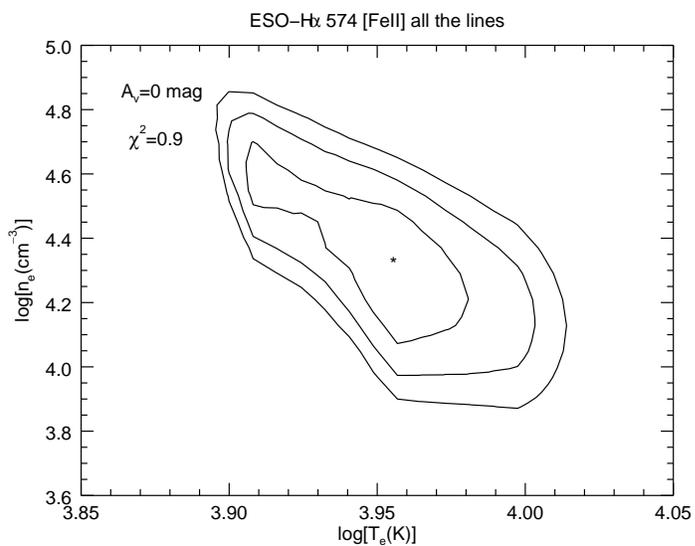}
\caption{\label{fig:chi_eso} $\chi^2$-contours  of the fit through the [\feii] lines  detected in ESO-H$\alpha$ 574. 
The curves refer to increasing values of $\chi^2$ of 30\,\%,  60\,\%, 90\,\%. The minimum $\chi^2$ value is given, as well.}
\end{figure}

\begin{figure}
\includegraphics[width=10cm]{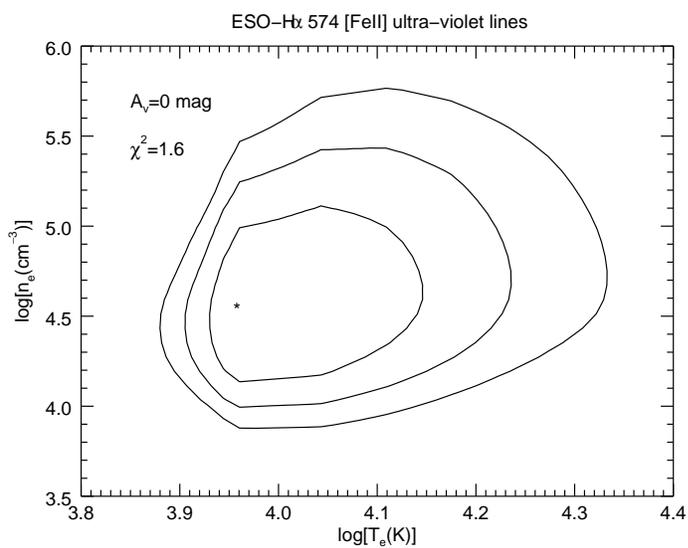}
\caption{\label{fig:chi_uv} As in Figure\,\ref{fig:chi_eso} for the fit of the [\feii] ultra-violet lines detected in ESO-H$\alpha$574.}
\end{figure}

\begin{figure}
\includegraphics[width=15cm]{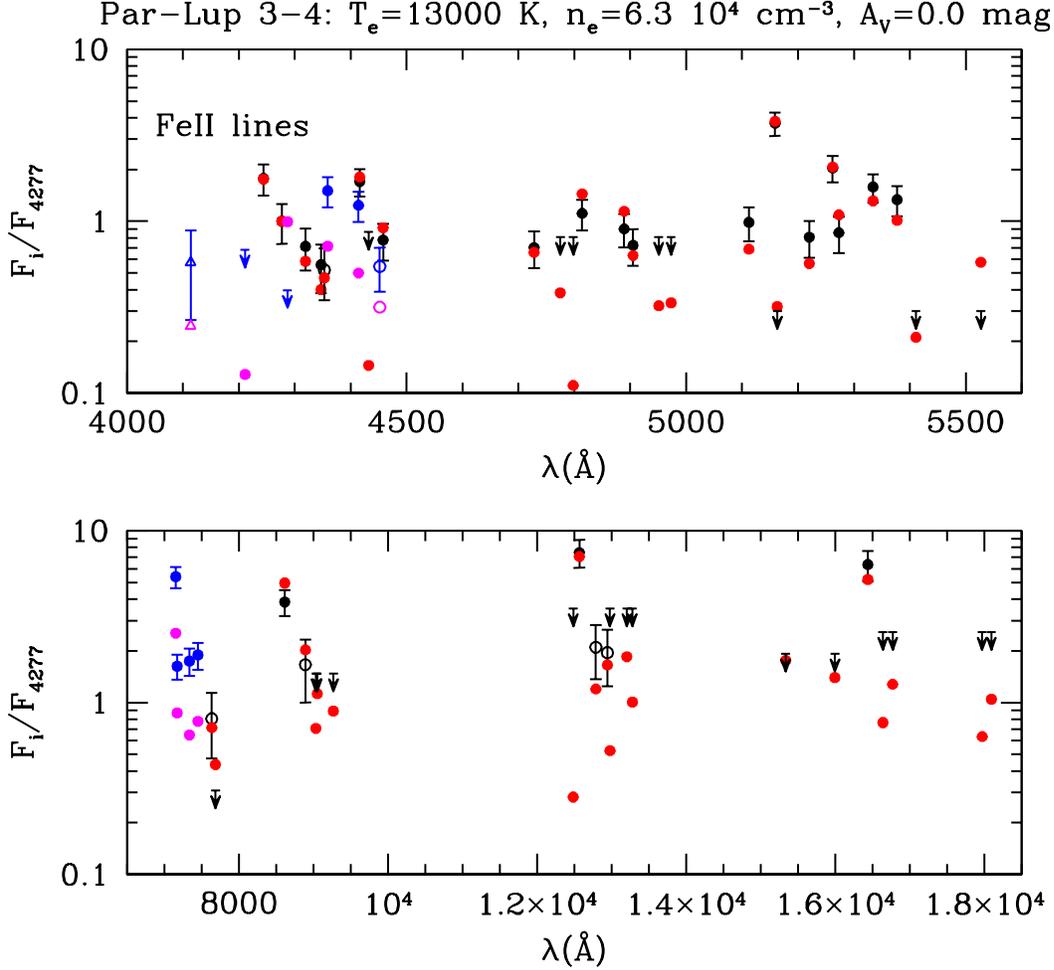}
\caption{\label{fig:bestfit_parlup}  NLTE best-fit model of the [\feii] lines  detected in Par-Lup 3-4. In the fitting procedure the lines detected with snr $\ge$ 5, represented as filled circles (black: data, red: model) have been included. Lines detected with 3\,$\le$ snr $<$\,5 and 2\,$\le$ snr $<$\,3 are reported with open circles and open triangles, respectively. Down arrows are the blended lines or the 2-$\sigma$ upper limits of lines not detected in Par-Lup 3-4 but detected in ESO-H$\alpha$ 574.
With blue and magenta symbols we indicate the observed data and model predictions of lines coming from levels b$^2$H, a$^6$S, and a$^2$G, which are not included in the fitting procedure. The best-fit parameters are reported as well.}
\end{figure}
\begin{figure}
\includegraphics[width=10cm]{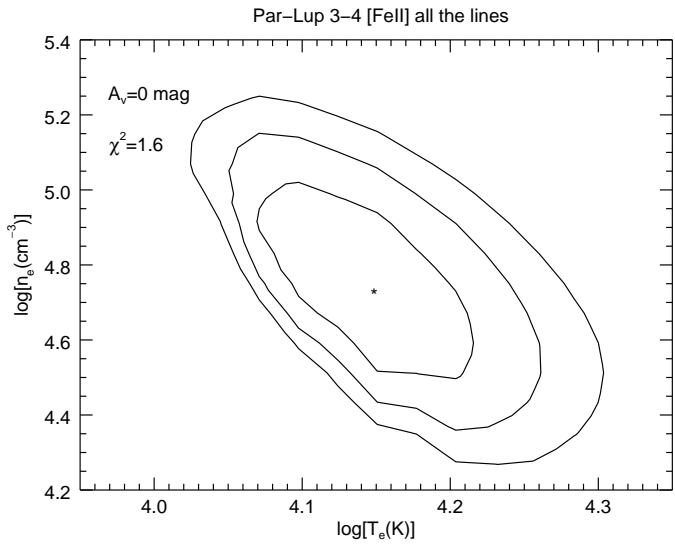}
\caption{\label{fig:chi_par} As in Figure\,\ref{fig:chi_eso} for the fit of the [\feii] lines detected in Par-Lup 3-4.}
\end{figure}

\begin{figure*}
\includegraphics[width=15cm]{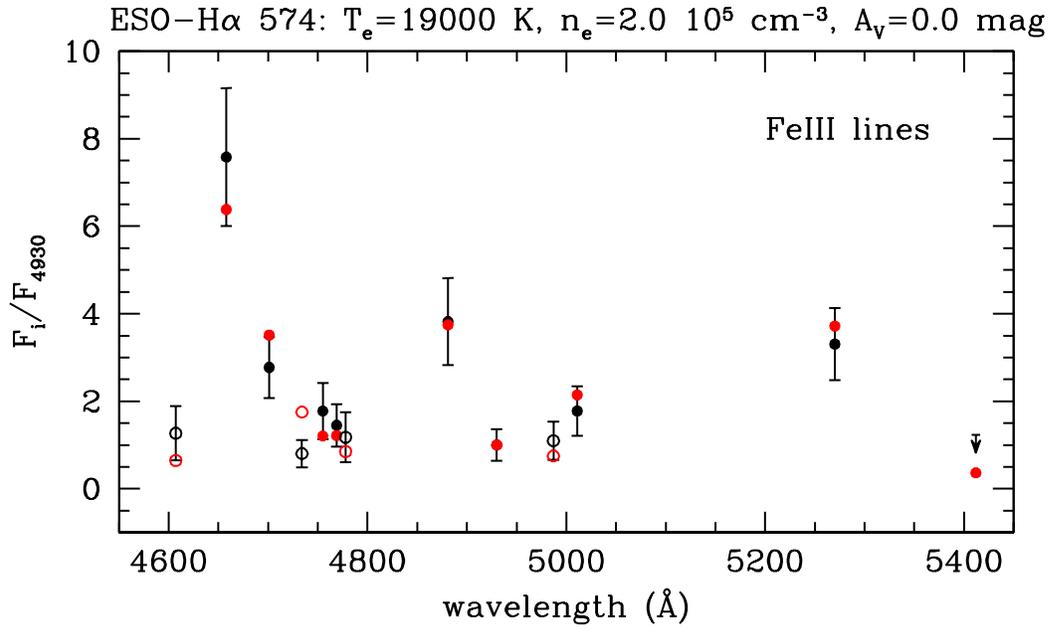}
\caption{\label{fig:fit_feiii} NLTE best-fit model of the [\feiii] lines  detected in ESO-H$\alpha$ 574. The symbols have the same meaning as in Figure\,\ref{fig:bestfit_eso}.}
\end{figure*}
\begin{figure}
\includegraphics[width=10cm]{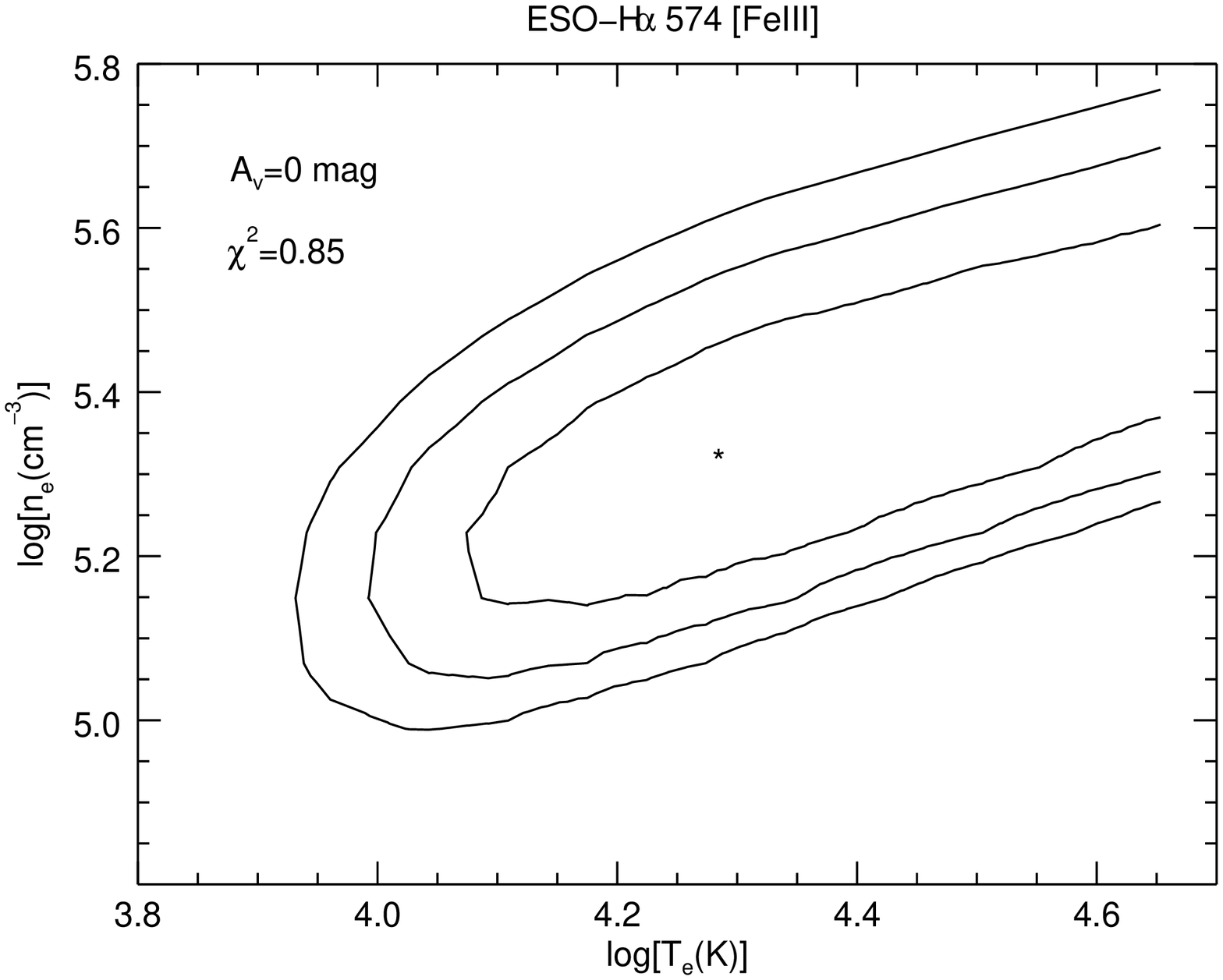}
\caption{\label{fig:chi_eso_feIII} As in Figure\,\ref{fig:chi_eso} for the fit of the [\feiii] lines detected in ESO-H$\alpha$ 574.}
\end{figure}

\begin{figure}
\includegraphics[width=12cm]{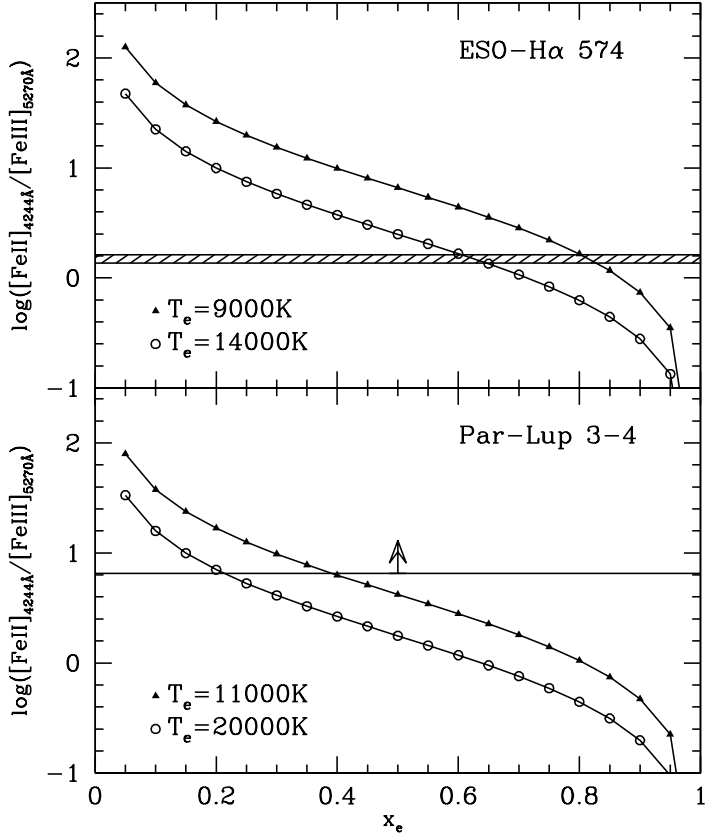}
\caption{\label{fig:fit_xe} [\feii]4244\AA/[\feiii]5270\AA\, line ratio as a function of the fractional ionization for different values of the electron temperature. The ratio measured in ESO-H$\alpha$ 574
(upper panel) and Par-Lup 3-4 (lower panel) is depicted with an horizontal line.}
\end{figure}

\begin{figure}
\includegraphics[width=13cm]{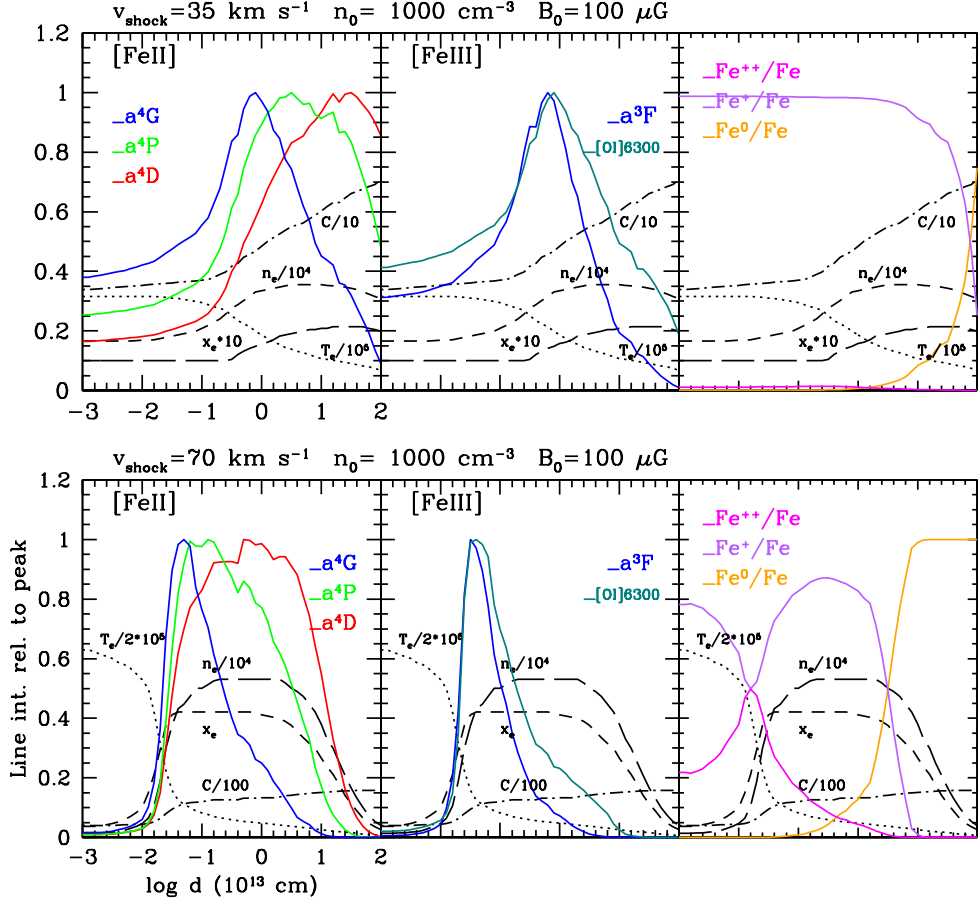}
\caption{\label{fig:fit_hart} Post-shock intensities relative to peak values vs. distance from the shock front, adapted from Figure\,1 of Hartigan, Morse, \& Raymond  (1994).  [\feii] and [\feiii] lines (left and middle panels, respectively) 
are shown for two shock velocities (35 km s$^{-1}$, upper panel, and 70 km s$^{-1}$, lower panel). For [\feii] lines, the peak-normalized intensity profile of ultra-violet (blue), optical (green) and near-infrared (red) lines is shown. Temperature (in K, divided by 10$^5$ for v$_{shock}$= 35 km s$^{-1}$, and 2 10$^5$ for v$_{shock}$= 70 km s$^{-1}$), electron density (in cm$^{-3}$, divided by 10$^4$), fractional ionization (multiplied by 10 for v$_{shock}$= 35 km s$^{-1}$), and compression factor ($C$ = $n_{\rm{post-shock}}/n_{\rm{pre-shock}}$, divided by 100 for v$_{shock}$= 70 km s$^{-1}$ and by 10 for v$_{shock}$= 35 km s$^{-1}$) are plotted with dotted, short-dashed, long-dashed, and dot-short-dashed curves, respectively. The assumed pre-shock gas conditions in terms of density and magnetic field strength are reported, as well. In the middle panels is also shown the peak-normalized intensity profile of the [\oi]\,6300 \,\AA\, line. The right panels give the relative fraction of Fe$^0$, Fe$^+$, and  Fe$^{++}$ with respect to the total Fe abundance along the shock profile. }
\end{figure}

%

\end{document}